\documentclass[12pt,leqno]{article}
\usepackage{latexsym}
\usepackage{amsfonts}
\usepackage{amsmath}
\usepackage{amsthm}

\def\tensor{\otimes}
\def\d{\partial}
\def\defpar{\vartheta}
\def\GF{{\bf F}}

\def\U{\mathcal U}
\def\g{h}
\def\id{\mathbf 1}

\newcommand{\eff}{{\rm eff}}
\newcommand{\st}[2]{\stackrel{#1}{#2}}

\newcommand{\dl}[1]{\displaystyle\frac{{\d}}{\d #1}}

\newcommand{\dd}[2]{\frac{\d #1}{\d #2}}

\newcommand{\ddl}[2]{\frac{\d^L #2}{\d #1}}

\newcommand{\vddl}[2]{{\frac{\delta #1}{\delta #2}}}

\newcommand{\ab}[2]{\big(#1,\,#2\big)}

\newcommand{\STAR}{*}

\newcommand{\qcommut}[2]{{[#1\stackrel{\STAR}{,}#2]}}
\newcommand{\commut}[2]{[#1,#2]}
\newcommand{\scommut}[2]{\{#1\stackrel{\STAR}{,}#2\}}
\newcommand{\cscommut}[2]{\{#1,#2\}}

\newcommand{\tr}{{\rm \,Tr}\,}
\def\half{{\frac{1}{2}}}
\newcommand{\p}[1]{|#1|}
\newcommand{\lied}{\mathcal L}

\newcommand{\bref}[1]{{\rm {\bf\ref{#1}}}}

\def\be{\begin{eqnarray}}
\def\ee{\end{eqnarray}}
\def\beann{\begin{eqnarray*}}
\def\eeann{\end{eqnarray*}}
\def\beq{\begin{equation}}
\def\eeq{\end{equation}}
\def\ba{\begin{array}}
\def\ea{\end{array}}
\def\ben{\begin{enumerate}}
\def\een{\end{enumerate}}
\def\bea{\begin{eqnarray}}
\def\eea{\end{eqnarray}}
\def\beann{\begin{eqnarray*}}
\def\eeann{\end{eqnarray*}}
\def\beq{\begin{equation}}
\def\eeq{\end{equation}}
\def\ba{\begin{array}}
\def\ea{\end{array}}
\def\ben{\begin{enumerate}}
\def\een{\end{enumerate}}

\def\5{\bar }
\def\6{\partial }
\def\7{\hat }
\def\4{\tilde }

\def\cU{{\cal U}}

\def\s0#1#2{\mbox{\small{$\frac{#1}{#2}$}}}

\def\f#1#2#3{{f_{#1#2}}^{#3}}

\def\lieg{\mathfrak g}

\newtheorem{theorem}{Theorem}

\newtheorem{prop}{Proposition}
\def\qed{\hbox{${\vcenter{\vbox{
\hrule height 0.4pt\hbox{\vrule width 0.4pt height 6pt
\kern5pt\vrule width 0.4pt}\hrule height 0.4pt}}}$}}

\newcommand{\hgz}{\7\gamma^{[0]}}
\newcommand{\hgo}{\7\gamma^{[1]}}

\numberwithin{equation}{section}

\begin{document}

\begin{titlepage}



\vspace{1.5cm}
\begin{centering}

\vspace{0.5cm}

{\bf{\Large Local BRST cohomology and Seiberg-Witten maps in
noncommutative Yang-Mills theory}}

\vspace{1.5cm}

{\large Glenn Barnich$^{*,a}$, Friedemann Brandt$^b$, and Maxim
Grigoriev$^{\dag,a,c}$}

\vspace{1.5cm}

$^a$Physique Th\'eorique et Math\'ematique,\\ Universit\'e Libre
de Bruxelles,\\
Campus Plaine C.P. 231, B--1050 Bruxelles, Belgium

\vspace{.5cm}

$^b$Max-Planck-Institute for Mathematics in the Sciences,\\
Inselstra\ss e 22--26, D--04103 Leipzig, Germany

\vspace{.5cm}

$^c$Tamm Theory Department, Lebedev Physical Institute,\\
Leninsky prospect 53, 119991 Moscow, Russia

\vspace{1.5cm}

\end{centering}
\vspace{.5cm}

\begin{abstract}
We analyze in detail the recursive construction of the
Seiberg-Witten map and give an exhaustive description of its
ambiguities. The local BRST cohomology for noncommutative
Yang-Mills theory is investigated in the framework of the
effective commutative Yang-Mills type theory. In particular, we
show how some of the conformal symmetries get obstructed by the
noncommutative deformation.
\end{abstract}

\vfill

\noindent \footnotesize{$^*$ Research Associate of the National
Fund for
  Scientific Research (Belgium).\\
  $^\dag$ Postdoctoral Visitor of the National Fund for Scientific Research (Belgium).}

\end{titlepage}

\tableofcontents

\vfill
\pagebreak

\section{Introduction}

Using arguments from string theory, noncommutative Yang-Mills
theory has been shown \cite{Seiberg:1999vs} to be equivalent to a
Yang-Mills type theory with standard gauge symmetries and an
effective action containing, besides the usual Yang-Mills term,
higher dimensional gauge invariant interactions. This equivalence
is implemented through a so-called Seiberg-Witten map (SW map), a
redefinition of both the gauge potentials and the parameters of
the gauge transformations.

By considering an expansion in some parameter of noncommutativity
$\defpar$, noncommutative Yang-Mills theory can be understood as a
consistent deformation of standard Yang-Mills theory in the sense
that the action and gauge transformations are deformed
simultaneoulsy in such a way that the deformed action is invariant
under the deformed gauge transformations. An appropriate framework
for analyzing such consistent deformations of gauge theories has
been shown \cite{Barnich:1993vg,Henneaux:1997bm} to be the
antifield-antibracket formalism (see
\cite{Becchi:1975nq,Tyutin:1975qk,Zinn-Justin:1974mc,Dixon:1975si}
in the Yang-Mills context,
\cite{Batalin:1981jr,Batalin:1983wj,Batalin:1983jr,Batalin:1984ss,%
Batalin:1985qj} for the generic case and
\cite{Henneaux:1992ig,Gomis:1995he} for reviews).

By reformulating the question of existence of SW maps in this
context, the whole power of the theory of ``anti'' canonical
transformations is available. In the generic case, this leads to
an ``open'' version of the gauge equivalence condition, valid only
up to terms vanishing when the equations of motion
hold~\cite{Barnich:2001mc}. These features have been shown to be
crucial for the construction of a SW map for the noncommutative
Freedman-Townsend model.

In the case of Yang-Mills and Chern-Simons theory, the relatively
simple structure of the gauge algebra allows one to analyze the SW
map using antifield independent BRST techniques
\cite{Okuyama:2001sw,Brace:2001fj,Brace:2001rd,Picariello:2001mu,%
Cerchiai:2002ss}. Nevertheless, from the point of view of
consistent deformations of gauge theories, it is sometimes useful
to lift these maps to anticanonical transformations in the
field-antifield space, as first discussed
in~\cite{Gomis:2000sp,Brandt:2001aa}.

The first objective of this paper is to improve the explicit
recursive construction of the SW map by cohomological methods. The
advantage of our solution is the use of explicit expressions for
the contracting homotopies adapted respectively to an expansion in
the deformation parameter and to an expansion in homogeneity of
the fields. These homotopies are based on previous works on the
local BRST cohomology of Yang-Mills theory
\cite{Brandt:1990gy,Dubois-Violette:1992ye}. The methods also
allow one to derive the general recursive solution to the SW gauge
equivalence condition. In particular, this solution contains
additional ambiguities besides those already discussed in
\cite{Asakawa:1999cu,Bichl:2001cq,Brace:2001rd}. We also point out
that the ambiguities in the SW map can be understood entirely in
the context of the standard Yang-Mills gauge field.

Another aim of this paper is to analyze the local, antifield
dependent BRST cohomology groups of noncommutative Yang-Mills
theory. These cohomology groups contain the information about the
(potential) anomalies, counterterms, and the global symmetries of
the model (see e.g.~\cite{Barnich:2000zw} and references therein).
In the space of formal power series in the deformation parameter,
the analysis can be done in terms of either the noncommutative or
the commutative formulation. It turns out to be more convenient to
work in the commutative formulation, because one can rely on known
results in standard Yang-Mills theory, adapted to the effective
theory with higher dimensional interactions. Representatives of
the cohomology classes in the noncommutative formulation can then
be obtained by applying the inverse SW map.

In the next section, we first recall the central equations for the
reformulation of SW maps in the antifield formalism. Then we make
some general observations on the relation between the BRST
cohomology groups of Yang-Mills theory and its noncommutative
deformation.

In the first part of section \bref{s3}, we show in detail that
existence of the SW maps follows a priori from known results on
the local BRST cohomology of standard Yang-Mills theory
\cite{Barnich:2000zw,Barnich:1994ve,Barnich:1995mt}. Even though
this kind of reasoning does of course not lead to new results for
noncommutative Yang-Mills theory, where the existence of SW maps
had been proved constructively in the original paper
\cite{Seiberg:1999vs}, it has been shown to be useful for the
existence proof of such maps in more complicated models
\cite{Barnich:2001mc}. In the next part we construct the
infinitesimal generating functional for the SW map understood as
an anticanonical transformation. The associated evolution
equations in the deformation parameter reproduce the original
differential equations from~\cite{Seiberg:1999vs}. We then give
recursive constructions of particular solutions for the SW
equations and analyze in detail the ambiguities in the general
solution. Finally, we address the question of triviality of the
whole noncommutative deformation. In particular, we give the
precise argument why noncommutative Yang-Mills theory is a non
trivial deformation. Then, we show from cohomological arguments
that noncommutative Chern-Simons theory is a trivial deformation
of its commutative version, as shown first in
\cite{Grandi:2000av}.

Section \bref{sec:coh} is devoted to the local BRST cohomology
of noncommutative Yang-Mills theory. Using a Seiberg-Witten map
the analysis is done in the commutative formulation. We show that
cohomology classes that do not involve the dynamics are unaffected
by the noncommutative deformation, while the others can be
obstructed. In particular, the breaking of the global Lorentz
invariance is discussed in some detail.

Finally, various more technical proofs are given in the
appendices.

\section{Generalities}
\label{sec:s2}

We assume the space-time manifold to be ${\mathbb R}^n$ with
coordinates $x^\mu\,,\mu=1,\dots,n$. Throughout the paper we use
notations and conventions from~\cite{Barnich:2001mc}. In
particular, the Weyl-Moyal star-product is defined through
\begin{equation}
 f*g
(x)=\exp\left(i{\wedge_{12}}\right) f(x_1)g(x_2)|_{x_1=x_2=x},\ \
\ \wedge_{12}= \frac{\defpar}{2}{\theta^{\mu\nu}}
\partial_\mu^{x_1}\partial_\nu^{x_2},
\end{equation}
for a real, constant, antisymmetric matrix $\theta^{\mu\nu}$. The
parameter $\defpar$ has mass dimension $-2$.

A natural space in deformation quantization is the space of formal
power series in $\defpar$ with coefficients in smooth functions.
In the context of local field theories, smooth functions are
replaced by local functions, i.e., functions that depend on
$x^\mu$, the fields, and a finite number of their derivatives.
More precisely, even though the whole series can depend on an
infinite number of derivatives of the fields, each monomial in
$\defpar$ involves only a finite number of them.

A noncommutative gauge theory is a consistent deformation of its
commutative counterpart, in the sense that the action and the
gauge transformations are simultaneously deformed in a compatible
way. An appropriate framework to describe such consistent
deformations is the Batalin-Vilkovisky formalism:
the deformations can be described entirely in terms of the master
action since it encodes both the gauge invariant action and
the gauge transformations.
Furthermore, the associated (antifield dependent) BRST
cohomology of the undeformed theory controls the deformation
\cite{Barnich:1993vg,Henneaux:1997bm}.

Consider then a noncommutative gauge theory described by the
minimal proper solution $\hat S[\hat\phi,\hat \phi^*;\defpar]$ of
the master equation,
\begin{equation}
\hat S[\hat\phi,\hat \phi^*;\defpar] =\sum_{s=0}^\infty \defpar^s
S^{(s)}[\hat\phi,\hat\phi^*]\,.
\end{equation}
In particular, the undeformed (commutative) theory is determined
by the master action
\begin{equation}
S^{(0)}[\hat\phi,\hat\phi^*] = \hat S[\hat\phi,\hat
\phi^*;\defpar]\Big|_{\defpar=0}\,.
\end{equation}

\subsection{Reformulation of SW maps in the
BV formalism}

A SW map is by definition a simultaneous field and gauge parameter
redefinition such that the gauge structure of the deformed theory
is mapped to that of the undeformed one. In the context of the
antifield formalism, the existence of a SW map can be expressed in
four equivalent ways.
\begin{enumerate}
\item \label{1stfor} There exists an anticanonical
field-antifield transformation\footnote{ Only anticanonical
transformation that reduce to the identity to order $0$ in the
deformation parameter are considered here. Invertibility of these
transformations in the space of formal power series is then
guaranteed.} $\hat\phi[\phi,\phi^*;\defpar],
\hat\phi^*[\phi,\phi^*;\defpar]$ such that
\begin{equation}
\hat
S[\hat\phi[\phi,\phi^*;\defpar],\hat\phi^*[\phi,\phi^*;\defpar];
\defpar]=
S^{\rm eff}_0[\phi;\defpar]+ \sum_{r\geq 1}
S^{(0)}_r[\phi,\phi^*]\,, \label{3.35}
\end{equation}
where $S^{\rm eff}_0[\hat\phi;0]=S^{(0)}_0[\hat\phi;0]$ and the
subscript denotes the antifield number.

\item There exists a generating functional of ``second type''
$\GF[\phi,\hat\phi^*;\defpar]$ with
\begin{equation}
\hat\phi^A(x) =\frac{\delta^L \GF}{\delta\hat\phi^*_A(x)}\,,
\qquad \phi^*_A(x) =\frac{\delta^L \GF}{\delta\hat\phi^A(x)}\,,
\label{transfo}
\end{equation}
such that
\begin{equation}
\hat S[\frac{\delta^L \GF}{\delta\hat\phi^*},\hat\phi^*;\defpar]= S^{\rm
eff}_0[\phi;\defpar]+\sum_{k\geq 1} S^{(0)}_k[\phi,\frac{\delta^L
\GF}{\delta \phi}]\,, \label{3.38}
\end{equation}
with initial condition $\GF=\int d^nx\
\hat\phi^*_A\phi^A+O(\defpar)$.

\item There exists a functional $\hat
  \Xi[\hat\phi,\hat\phi_*;\defpar]$ such that
\begin{equation}
 \frac{\partial \hat S}{\partial \defpar}
 =\hat B_0+(\hat S,\hat \Xi) \label{eq:diff-cond}
\end{equation}
holds with $\hat B_0[\hat\phi\,;\defpar]$. The field-antifield
redefinition of formulation \ref{1stfor} can then be constructed
as the solution to the differential equations \bea
    \frac{\partial \hat\phi^A}{\partial \defpar}&=&
    \ab{\hat \Xi}{\hat\phi^A},\label{canon}\\
    \frac{\partial \hat \phi^*_A}{\partial \defpar}&=&
    \ab{\hat \Xi}{\hat\phi^*_A}\,\label{canon*}
\eea and $\hat B_0[\hat\phi\,;\defpar] =({\partial S^{\rm
eff}_0}/{\partial
\defpar})[\phi[\hat\phi\,;\defpar];\defpar]$.
Formally, this solution can be written as
\begin{equation}
  \begin{aligned}
\hat\phi^A(x)&=[P\exp\int_0^\defpar d\defpar^\prime
(\Xi(\defpar^\prime),\cdot)]\phi^A(x),\\
\hat\phi^*_A(x)&=[P\exp\int_0^\defpar d\defpar^\prime (
\Xi(\defpar^\prime),\cdot)]\phi^*_A(x)\,, \label{3.311}
\end{aligned}
\end{equation}
where $\Xi[\phi,\phi^*;\defpar]$ is the same function of
$\phi,\phi^*;\defpar$ as $\hat\Xi[\hat\phi,\hat\phi^*;\defpar]$ is
of $\hat\phi,\hat\phi^*;\defpar$.

\item The deformed and undeformed theories are weakly gauge equivalent
  in the following sense. Let $\hat L_0[\hat\varphi;\defpar]$
be a Lagrangian of the deformed theory and $L_0^{\rm
eff}[\varphi;\defpar]=\hat L_0[f[\varphi;\defpar]]$ be the
respective effective Lagrangian. In the case of an irreducible
gauge theory, there exists a simultaneous redefinition of the
original gauge fields $\hat\varphi^i=f^i[\varphi;\defpar]$ and the
parameters\footnote{A square bracket means a local dependence on
the fields and their derivatives, while the round bracket means
that this dependence is linear and homogeneous.}
$\hat\epsilon^\alpha=g^\alpha_\beta[\varphi;\defpar](\epsilon^\beta)$
of the irreducible generating set of nontrivial gauge
transformations $\hat
R^i_\alpha[\hat\phi;\defpar](\hat\epsilon^\alpha)$ such that
\begin{equation}
(\hat\delta_{\hat\epsilon}
\hat\varphi^i\big)|_{\hat\varphi=f,\hat\epsilon=g} \approx
\delta_\epsilon f^i\,,
\end{equation}
where $\approx$ means terms that vanish when the equations of
motions associated to $S^{\rm eff}_0=\int d^nx\  L^{\rm
eff}_0[\varphi;\defpar]$ hold. The operators
$\hat\delta_{\hat\epsilon}, \delta_\epsilon$ are given by
\begin{align}
  \hat\delta_{\hat\epsilon}&=\sum_{k=0}\partial_{\mu_1}\dots\partial_{\mu_k}
  \Big(\hat
  R^i_\alpha[\hat\varphi;\defpar](\hat\epsilon^\alpha)\Big)\frac{\partial}{\partial
    (\partial_{\mu_1}\dots\partial_{\mu_k} \hat\varphi^i)},\\
  \delta_\epsilon&=\sum_{k=0}\partial_{\mu_1}\dots\partial_{\mu_k}\Big(
  R^i_\alpha[\varphi](\epsilon^\alpha)\Big)\frac{\partial}{\partial
    (\partial_{\mu_1}\dots\partial_{\mu_k} \varphi^i)}\,,
\end{align}
with $R^i_\alpha[\varphi](\epsilon^\alpha)$ being the associated
irreducible generating set of nontrivial gauge transformations of
the undeformed theory.
\end{enumerate}

In what follows, by effective theory we mean the commutative
theory described by the Lagrangian $L^{\rm
eff}_0[\varphi;\defpar]$ or, equivalently, by the solution of the
master equation given by the right hand side of \eqref{3.35}.

\subsection{Local BRST cohomology and SW maps}

Let $\hat s=(\hat S,\cdot)$ be the BRST differential of a
noncommutative theory admitting a SW map. In the space of formal
power series in the deformation parameter with coefficients in
local functions or local functionals, the BRST cohomology groups
$H(\hat s)$ are isomorphic to the BRST cohomology groups of the
associated effective theory because these two theories are related
by an anticanonical field-antifield redefinition. Furthermore,
these cohomology groups are included in the associated cohomology
groups of the undeformed commutative theory, evaluated in the
spaces of formal power series in $\defpar$.

The BRST differential $s^{\rm eff}$ of the effective theory can be
expanded according to the antifield number as $s^{\rm
eff}=\delta^{\rm eff} +\gamma+s_1+\dots$. The Koszul-Tate
differential $\delta^{\rm eff}$ lowers the antifield number by $1$
and is associated to the equations of motion of $S^{\rm
eff}_0[\varphi;\defpar]$, while $\gamma, s_1,\dots$ of antifield
number $0, 1, ...$ are identical to the corresponding operators of
the undeformed theory.

In the Yang-Mills case that we are interested in here, $\gamma$ is
a differential and the operators $s_1,\dots$ all vanish. Hence,
the difference between the local BRST cohomology groups of
noncommutative Yang-Mills theory and its commutative counterpart
is due only to the dynamics encoded in $\delta^{\rm eff}$
respectively $\delta$.

\section{SW maps in noncommutative Yang-Mills theory}
\label{s3}

In the first subsection, we give the master action for
noncommutative Yang-Mills theory and the associated BRST
differential involving the antifields. We assume that fields take
values in $u(N)$ or in some associative matrix algebra $\cU$. For
simplicity we limit ourselves to pure Yang-Mills theories. The
inclusion of matter fields is straightforward along the lines of
\cite{Barnich:2002pb}.

\subsection{Master action and BRST differential}

The minimal (not necessarily proper) solution of the master
equation for noncommutative Yang-Mills theory is given by
\begin{equation}
\hat S=\int d^nx\ \Big( -\frac{1}{4\kappa^2} {\rm Tr}\ (\hat
F^{\mu\nu}\STAR\hat F_{\mu\nu}) +\hat A^{*\mu}_A\STAR (\hat D_\mu
\hat C)^A +\hat C^*_A\STAR(\hat C\STAR\hat C)^A\Big)\,,
\label{eq:YM-maction}
\end{equation}
where $\hat A_\mu=\hat A_\mu^A T_A$ and $\hat C=\hat C^A T_A$ are
either $u(N)$ or $\cU$-valued gauge fields and ghost fields, $\hat
A^{*\mu}_A$ and $\hat C^{*}_A$ are the antifields conjugate to
$\hat A_\mu^A$ and $\hat C^A$, and
\[
\hat F_{\mu\nu}=\6_\mu \hat A_\nu-\6_\nu \hat A_\mu +\qcommut{\hat
A_\mu}{\hat A_\nu}, \quad \hat D_\mu \hat C=\6_\mu \hat
C+\qcommut{\hat A_\mu}{\hat C},
\]
with $\qcommut{\ }{\ }$ the graded star-commutator,
\begin{equation}
  \qcommut{A}{B}=A\STAR B-(-1)^{\p{A}\p{B}} B\STAR A\,.
\end{equation}

The BRST differential $\hat s$ for the noncommutative model is
defined in the standard way as canonically generated by the
associated master action: \bea \hat s\,\cdot\,=\ab{\hat S}{\,
\cdot\, }. \label{hats} \eea It is useful to represent $\hat s$ as
the sum $\hat s = \hat \gamma + \hat \delta$, where $\hat\gamma$
is the part of the BRST differential with antifield number $0$ and
$\hat \delta$ is the Koszul-Tate part (see e.g.
\cite{Henneaux:1992ig}). The differentials $\hat \gamma$ and $\hat
\delta$ act on the gauge fields, ghost fields, and antifields as
follows:
\begin{equation}
\begin{aligned}
\label{eq:gamma-delta}
&\begin{aligned}
  \hat \gamma \hat A_\mu &= \hat D_\mu \hat C, &\qquad \hat \gamma
  \hat C &= - \hat C\STAR\hat C,
  \\
  \hat \delta \hat A_\mu &= 0,&\qquad \hat \delta \hat C&=0\,,
\end{aligned}
\\
&\begin{aligned}
  \hat \gamma \hat A^{*\mu}_A &= -(T_A \7C)^B
  \STAR\7A^{*\mu}_B-\7A^{*\mu}_B\STAR (\7C T_A)^B,
  \\
  \hat \gamma \hat C^*_A &= -(T_A \7C)^B \STAR\7C^*_B+\7C^*_B\STAR
  (\7C T_A)^B,
  \\
  \hat \delta \hat A^{*\mu}_A &=\displaystyle{\frac{1}{\kappa^2}} {\rm
    Tr}\,(T_A \hat D_\nu \hat F^{\nu\mu}),
  \\
  \hat \delta \hat C^*_A &= - \6_\mu \hat A^{* \mu}_A -(T_A \7A_\mu)^B
  \STAR\7A^{*\mu}_B+\7A^{*\mu}_B\STAR (\7A_\mu T_A)^B,
  \end{aligned}
\end{aligned}
\end{equation}
where $T_A\7C=(T_A \7C)^BT_B$. The BRST differential $\hat s$ can
be expanded in $\defpar$,


In the next subsection, we show that existence of the SW map
follows directly from standard results on the BRST cohomology of
the commutative Yang-Mills theory.

\subsection{Existence of SW map from BRST cohomology}

As discussed in \cite{Barnich:2001mc}, existence would be direct
if there were no antifield dependent cohomology, but
even in the $u(N)$ case
there is in fact antifield dependent
cohomology because
of the U(1) factor.
We show that the noncommutative deformation does not
involve this cohomology, which completes the previous arguments.

Assume we have constructed a SW map
$\hat\phi^{k}[\phi,\phi^*;\defpar],\hat\phi^{*k}[\phi,\phi^*;\defpar]$
to order $k$ in $\defpar$. This means that~\eqref{3.35} holds up
to terms of order $k+1$ and higher:
\begin{multline}
\hat S
[\hat\phi^{k}[\phi,\phi^*;\defpar],\hat\phi^{*k}[\phi,\phi^*;\defpar];\defpar]
=
\\
= \sum_{l=0}^k \defpar^l S_0^{{\rm eff}\,(l)}[A] +\sum_{r\geq
1}S^{(0)}_r[\phi,\phi^*] +\defpar^{k+1}\,\tilde
S^{(k+1)}[\phi,\phi^*]+O(k+2)\,, \label{fb2}
\end{multline}
with $\ab{S^{(0)}}{S_0^{{\rm eff}\,(l)}[A]}=0$ and
$\hat\phi^{k},\hat\phi^{*k}$ related to $\phi,\phi^*$ through an
anticanonical transformation. The master
equation $\ab{\hat S}{\hat S}=0$ then implies
\begin{equation}
  \ab{S^{(0)}}{\tilde S^{(k+1)}}=0\,.\label{coc0}
\end{equation}
Suppose that the cocycle ${\tilde S^{(k+1)}}$ belongs to a
subspace $\mathfrak S$, where the representatives of the
cohomology of $s^{(0)}$ can be chosen to be antifield independent,
so that
\begin{equation}
\label{eq:S-represent}\ab{S^{(0)}}{\tilde S^{(k+1)}}=0\ \ \ \
\Longrightarrow\ \ \ \  \tilde S^{(k+1)}=S^{{\rm
eff}(k+1)}_0[A]+\ab{S^{(0)}}{\tilde\Xi^{(k+1)}}.
\end{equation}
Then, the SW map can be constructed as a succession of
anticanonical transformations.  Indeed, if we define
\begin{equation}
  \begin{aligned}
    \hat\phi^{k+1}[\phi,\phi^*;\defpar]&
    =\exp{}\left(\defpar^{k+1}\ab{\tilde\Xi^{(k+1)}}{\,\cdot\,}
    \right)\hat\phi^k\,,\\
    \hat\phi^{*k+1}[\phi,\phi^*;\defpar]&
    =\exp{}\left(\defpar^{k+1}\ab{\tilde\Xi^{(k+1)}}{\,\cdot\,}
    \right)\hat\phi^{*k}\,,
\end{aligned}
\end{equation}
the action $\hat S[\hat\phi^{k+1},\hat\phi^{*k+1};\defpar]$
satisfies~\eqref{fb2} with $k+1$ in place of $k$. For $k=0$,
Eq.~\eqref{fb2} obviously holds with $\hat\phi^{0}=\phi$,
$\hat\phi^{*0}=\phi^*$ the identity map and $S_0^{{\rm
eff}\,(0)}[A]$ the standard commutative Yang-Mills action.

In our case, the subspace $\mathfrak S$ can be taken to be the
space of local functionals depending at most linearly on
antifields and depending on the ghosts only via their derivatives
when written in terms of undifferentiated antifields. We show in
appendix~{\bf A} that (i) equation \eqref{eq:S-represent} indeed
holds if ${\tilde S^{(k+1)}}\in \mathfrak S$ and (ii) that
$\tilde\Xi^{(k+1)}$ can be chosen  in $\mathfrak S$. In this case
${\tilde S^{(k+2)}}$ can also be chosen in $\mathfrak S$ (through
integrations by parts) since all terms of $\hat S$ of first and higher
order belong to $\mathfrak S$, and terms in
$\hat\phi^{k+1}[\phi,\phi^*;\defpar]$, $
\hat\phi^{*k+1}[\phi,\phi^*;\defpar]$ of first and higher order are
at most linear in antifields and depend only on differentiated
ghosts if $\tilde\Xi^{(l)}$, $l=1\dots,k+1$ belong to $\mathfrak
S$, which completes the proof.

\subsection{Explicit construction of generating functional}\label{s4}

In this subsection, we give a constructive approach to the
differential equations of Seiberg and Witten, whose integration
provides the map that establishes the equivalence of the gauge
structure of the noncommutative and the commutative theories. The
differential equations appear here as those for an anticanonical
transformation, more precisely a Hamiltonian evolution equation
with time replaced by $\defpar$, the parameter of
noncommutativity. Because this generating functional contains the
evolution parameter $\defpar$, the formal solution is obtained by
the standard path-ordered exponential associated to time dependent
anticanonical transformation.

The generating functional $\hat\Xi$ defined by
\eqref{eq:diff-cond} \bea \dd{\hat S}{\defpar}=\hat B_0+\7s\,
\7\Xi\,, \label{full1} \eea can be constructed by using an
appropriate contracting homotopy. In order to do so, one
decomposes the functionals according to the antifield number and
expands in homogeneity in the fields. In this case the relevant
differential controling the construction is $\hgz$ which acts on
the fields and antifields simply according to \bea \hgz
\7A_\mu=\6_\mu\7C,\quad \hgz \7C=\hgz \7A^{*\mu}_A=\hgz
\7C^{*}_A=0\,. \label{B1b} \eea The decomposition of $\7\gamma$
then reads \bea \7\gamma=\hgz+\hgo. \label{B1a} \eea where $\hgo$
contains the quadratic terms of the $\7\gamma$-transformations.
Details are given in appendix~{\bf B} and we present here
  only the final results:
\begin{align}
    \7\Xi&=
  \frac{i\theta^{\alpha\beta}}{4}\int d^nx\ ( -\7A^{*
    \mu}_A\,\scommut{\7F_{\alpha\mu}+\d_\alpha \7A_\mu}{\7A_\beta}^A
  +\7C^{*}_A\,\scommut{\7A_\alpha}{\d_\beta \7C}^A),
\label{hatxi}\\
\hat B_0&=
\frac{i\theta^{\alpha\beta}}{\kappa^2}\int
d^nx\ {\rm Tr}(\frac 18 \7F_{\alpha\beta}\, \7F_{\mu\nu}\STAR
\7F^{\mu\nu} -\frac 12\7F_{\alpha\mu}\STAR \7F_{\beta\nu}\,
\7F^{\mu\nu}). \label{hateff}
\end{align}
The associated differential equations (\ref{canon}) and
(\ref{canon*}) for $\hat A_\mu$ and $\7C$ are the ones
from~\cite{Seiberg:1999vs}:
\begin{equation}
\begin{aligned}
\dd{\hat A_\mu}{\defpar}&= \ab{\7\Xi}{\hat A_\mu}=
-\frac{i\theta^{\alpha\beta}}{4}\,
\scommut{\7A_\alpha}{\7F_{\beta\mu}+\d_\beta \7A_\mu},
\\
\dd{\hat C}{\defpar}&= \ab{\7\Xi}{\hat C}=
\frac{i\theta^{\alpha\beta}}{4}\, \scommut{\d_\alpha
\7C}{\7A_\beta}. \label{SWdiff}
\end{aligned}
\end{equation}

Notice also that equation~\eqref{hateff} provides directly
the effective action to first order in $\defpar$:
\begin{multline}
 S^{\rm eff}[A;\defpar]=\\
= -\frac{1}{4 \kappa^2} \int d^nx \,{\rm Tr}\,
\left(F_{\mu\nu}F^{\mu\nu} + \frac{i\defpar
\theta^{\alpha\beta}}{2} ( - F_{\alpha\beta}F_{\mu\nu} F^{\mu\nu}
+4 F_{\alpha\mu} F_{\beta\nu} F^{\mu\nu}\label{3.22} )
\right)+O(\defpar^2)\,.
\end{multline}

\subsection{Recursive construction of SW map}

In this subsection we construct recursive solutions for the SW
map. We use the standard technique of homological perturbation
theory based either on an expansion in the deformation parameter
or an expansion in homogeneity in the fields. In the former case
the appropriate coboundary operator is $\bar \gamma=
\gamma+\commut{C}{\cdot\,}$ while in the latter it is
$\gamma^{[0]}$.

\subsubsection{Defining equations\label{subsec:def-eq}}
Linearity in antifields of the generating functional $\hat \Xi$
implies that the generating functional $\GF$ of second type can
also be chosen linear in antifields, \bea \GF=\int d^nx\
\left(\hat A^{*\mu}_A f^A_\mu+ \hat C^{*}_A\g^A\right)\,, \eea
where $f^A_\mu=f_\mu^A[A;\defpar]$ and $\g^A=\g^A[A;\defpar](C)$.
Then the defining equation~\eqref{3.38} reduces to
\begin{align}
   -\frac{1}{4\kappa^2}\int d^nx \tr \hat F_{\mu\nu}\hat F^{\mu\nu}
  [f;\defpar]=\int d^nx\ L^{\rm eff}_0[A;\defpar]\,,\\
  (\partial_\mu \g +[ f_\mu\stackrel{*}{,}\g])^A=\gamma f^A_\mu,\label{3.29}\\
 \frac{1}{2}\qcommut{\g}{\g}^A +\gamma \g^A=0\,,
\label{3.30}
\end{align}
where $f_\mu=f_\mu^A[A;\defpar]T_A$ and $\g=\g^A[A,C;\defpar]T_A$
and $\gamma$ is the gauge part of the BRST differential of the
commutative theory:
\begin{multline}
\label{eq:comm-gamma} \gamma f[A,C]=\sum_{k=0}\,\d_{\rho_1}\ldots
\d_{\rho_k}
  (D_\mu C)^B\ddl{(\d_{\rho_1}\ldots \d_{\rho_k} A^B_{\mu})}{f[A,C]}
\\
- \sum_{k=0}\,\frac{1}{2}\d_{\rho_1}\ldots
\d_{\rho_k}({f_{DE}}^BC^DC^E) \ddl{(\d_{\rho_1}\ldots \d_{\rho_k}
C^B)}{f[A,C]}\,,
\end{multline}
where $f_{DE}{}^B$ are the structure constants of the Lie algebra
associated to the underlying matrix algebra,
\begin{equation}
\label{Liealg}
[T_A,T_B]=f_{AB}{}^C T_C\,.
\end{equation}
Notice that for a general generating functional $\GF$ that is not
linear in the antifields, equations \eqref{3.29} and \eqref{3.30}
may contain equation of motion terms~\cite{Barnich:2001mc}.

Equation \eqref{3.29} is the SW equation (3.3) of
\cite{Seiberg:1999vs} under the form
\begin{equation}
\hat \delta_{\hat\lambda} \hat A=\delta_\lambda\hat A,\label{3.31}
\end{equation}
with the identifications $\hat A_\mu\leftrightarrow f_\mu$, $\hat
\lambda\leftrightarrow \g$ and $\lambda\leftrightarrow C$. We are
going to solve the BRST version~\eqref{3.30} of the integrability
condition before solving the SW equation \eqref{3.29}, because it
contains as unknown functions only the noncommutative gauge
parameter $\g$ as a function of $\theta^{\mu\nu}$, $C^A$,
$A_\mu^A$, and their derivatives.

\subsubsection{Expansion in $\defpar$}
Assume first that $\g^k=\sum_{l=0}^k
\defpar^{l}\g^{(l)}$ solves \eqref{3.30} up to order
$k+1$ in the deformation parameter~$\defpar$. This means that
\begin{equation}
\label{eq:k-order}
  \half \qcommut{\g^k}{\g^k}+\gamma \g^k=\defpar^{k+1}r^{(k+1)}+O(k+2)\,,
\end{equation}
where in the right hand side we have explicitly written the term of order
$k+1$. Applying $\gamma +\qcommut{\g^k}{\cdot}$ to this equation
the left hand side vanishes identically. The consistency condition
\begin{equation}
 ( \gamma +\qcommut{\g^k}{\cdot})(\defpar^{k+1}r^{(k+1)}+O(k+2))=0
\end{equation}
implies to lowest order
\begin{equation}
  (\gamma+\commut{C}{\cdot})r^{(k+1)}=0\,.
\end{equation}
Thus the operator $\bar\gamma=\gamma+\commut{C}{\cdot}$ is the
natural operator for recursively solving~\eqref{3.30} and
also~\eqref{3.29}\footnote{In this context the relevance of this
operator was first observed in~\cite{Brace:2001fj} (see
also~\cite{Picariello:2001mu}).}. In particular it is nilpotent,
$\bar\gamma^2=0$, when acting on matrix algebra-valued local
functions. Note that unlike the ordinary BRST differential
$\gamma$, $\bar\gamma$ doesn't commute with $\d_\mu$. Instead,
$\bar\gamma$ commutes with the covariant derivative:
$\commut{\bar\gamma}{D_\mu}=0$.

To proceed to the next order one has to show that the cocycle
$r^{(k+1)}$ is a coboundary of $\bar\gamma$. This can be achieved
by using  $\rho$ given explicitly by~\eqref{rho} which is a
contracting homotopy for $\bar\gamma$. Indeed, one can check that
\begin{equation}
\{\bar\gamma,\rho\} f(y,z,\chi,C)= f(y,z,\chi,C)-f(0,0,\chi,C)\,,
\end{equation}
where $f$ is an algebra-valued local function and the following
independent coordinates are introduced in the space of fields and
their derivatives (see appendix~{\bf A} for details):
\begin{equation}
\begin{gathered}
\{y^\alpha\}=\{\partial_{(\nu_1}\dots\partial_{\nu_l}
A^A_{\mu)}\}, \qquad
\{z^\alpha\}=\{\partial_{(\nu_1}\dots\partial_{\nu_l}D_{\mu)}  C^A\},\\
\{\chi^A_\Delta\} = \{D_{(\nu_1}\dots D_{\nu_{l}}
F^{{}A}_{\mu)\lambda}\}\,,\qquad \{C^A\}\,.\label{3.33}
\end{gathered}
\end{equation}
\begin{prop}\label{lemma:rec1}
A particular recursive solution $\g=\sum_{k=0}\defpar^k\g^{(k)}$
to equation \eqref{3.30} with $\g^{(0)}=C$ is given by
\begin{equation}
\label{eq:g-solution} \g^{(k+1)}=-\rho\left( \sum_{i+j+l=k+1}\half
\qcommut{\g^{(i)}}{\g^{(j)}}^{(l)}\right)\,, \quad 0 \leq i,j \leq
k,\,\,\, 0 \leq l \leq k+1, \quad k \geq 0\,,
\end{equation}
where we have expanded the $*$-commutator with respect to powers
in $\defpar$:
\begin{equation}
\qcommut{a}{b}=\sum_{l=0}^\infty \defpar^l \qcommut{a}{b}^l=
\commut{a}{b}+\frac{i\defpar \theta^{\alpha\beta}}{2} \{\d_\alpha
a,\d_\beta b\}+\cdots\,.
\end{equation}
\end{prop}
\begin{proof}
At zeroth order in $\defpar$ Eq.~\eqref{3.30} is satisfied by
$\g^{0}=C$. Assume that we have constructed $\g^{k}=\sum_{l=0}^k
\defpar^l \g^{(l)}$ such that $\g^{k}$ solves~\eqref{3.30} up to order
$k+1$ with $\g^{(l)}$ independent on undifferentiated ghosts for
$1 \leq l \leq k$.  At order $k+1$ in $\defpar$ Eq.~\eqref{3.30}
requires
\begin{equation}
\label{6.16} - \bar\gamma \g^{(k+1)} =
\frac{1}{2}\sum_{i+j+l=k+1}\qcommut{\g^{(i)}}{\g^{(j)}}^{(l)}\,,
\qquad 0 \leq i,j \leq k,\quad  0 \leq l \leq k+1\,.
\end{equation}
The right hand side of~\eqref{6.16} is just $r^{(k+1)}$
from~\eqref{eq:k-order} and therefore it is $\bar\gamma$-closed.
One can check that  $r^{(k+1)}$ does not depend on undifferentiated
ghosts. Indeed, among $\g^{(l)},\, 0 \leq l\leq k$, only
$\g^{(0)}$ depends on undifferentiated ghosts but it appears only
inside the $*$-commutator. One then finds that $\g^{(k+1)}=-\rho
r^{(k+1)}$ solves \eqref{6.16} because $r^{(k+1)}$ vanishes when
$y=z=0$. Furthermore, $\g^{(k+1)}$ is also independent on the
undifferentiated ghosts so that the construction can be iterated.
\end{proof}
In the case of $u(N)$-valued fields it follows from the recursive
construction that $\g$ also takes values in $u(N)$ because only
commutators or anticommutators  multiplied by imaginary unit are
involved.

\begin{prop}\label{lemma:f-solution}
For $\g=\sum_{k=0} \defpar^k \g^{(k)}$ as in
proposition~\bref{lemma:rec1} a particular recursive solution
$f_\mu=\sum_{k=0}\defpar^k f_\mu^{(k)}$ to equation \eqref{3.29}
with ${f_\mu}^{(0)}=A_\mu$ is given by
\begin{equation}
f_\mu^{(k+1)} = \rho \left( D_\mu \g^{(k+1)}
+\sum_{i+j+l=k+1}\qcommut{{f_\mu}^{(i)}}{\g^{(j)}}^{(l)}\right)\,,
\qquad  k\geq 0,
\end{equation}
with $0 \leq i,j \leq k$, $0 \leq l \leq k+1$.
\end{prop}
\begin{proof}
At zeroth order in $\defpar$ Eq.~\eqref{3.29} is satisfied by
$f^{0}_\mu=f^{(0)}_\mu=A_\mu$. Assume that we have constructed
$f_\mu^{k}=\sum_{l=0}^k
\defpar^l f^{(l)}_\mu$ that solves \eqref{3.29}
to order $k$, i.e.,
\begin{equation}
\gamma f_\mu^{k}-\d_\mu \g -\qcommut{f_\mu^{k}}{\g}=\defpar^{k+1}
t_\mu^{(k+1)}+O(k+2)\,.
\end{equation}
Applying $\gamma+\qcommut{\g}{\cdot}$ to both sides and
using~\eqref{3.30} one gets at order $k+1$ in $\defpar$ that
$\bar\gamma t_\mu^{(k+1)}=0$. Explicitly, $t_\mu^{(k+1)}$ is given
by
\begin{equation}
t_\mu^{(k+1)}=-D_\mu \g^{(k+1)}-\sum_{i+j+l=k+1}
\qcommut{{f_\mu}^{(i)}}{\g^{(j)}}^{(l)}\,, \qquad 0 \leq i,j \leq
k, \qquad  0 \leq l \leq k+1\,.
\end{equation}
At order $k+1$ in $\defpar$ Eq.~\eqref{3.29} requires:
\begin{equation}
\begin{aligned}
\label{eq:k+1} \bar\gamma f_\mu^{(k+1)} ~~=&~~ D_\mu
h^{(k+1)}+\sum_{i+j+l=k+1}\qcommut{{f_\mu}^{(i)}}{\g^{(j)}}^{(l)}~~=~~-t^{(k+1)}_\mu\,,
\\
& \qquad 0 \leq i,j \leq k,\quad 0 \leq l \leq k+1\,.
\end{aligned}
\end{equation}
Thus $f_\mu^{(k+1)}=-\rho t_\mu^{(k+1)}$ is a particular solution
to equation~\eqref{eq:k+1} because $t_\mu^{(k+1)}$ does not depend
on undifferentiated ghosts for reasons analogous to those
as in the previous proof.
\end{proof}
The same arguments as above show
that in the case of $u(N)$-valued fields $f_\mu$ is also
$u(N)$-valued.

\subsubsection{Expansion in homogeneity in the fields}

We now discuss a recursive solution for the SW map based on an
expansion according to the homogeneity in the fields. As in the
construction of the generating functional the relevant
differential is $\gamma^{[0]}$ corresponding to the Abelian theory
but now written in terms of unhatted variables. The associated
contracting homotopy $\rho^{[0]}$ is defined by~\eqref{A31} and
satisfies~\eqref{A333} with unhatted variables substituting for
hatted ones.
\begin{prop}\label{lemma:rec1-hom}
A particular recursive solution $\g=\sum_{k=1}\g^{[k]}$ to
equation \eqref{3.30} with $\g^{[1]}=C$ is given by
\begin{equation}
\g^{[k]}=-\rho^{[0]}(\gamma^{[1]} \g^{[k-1]}+
\sum_{l=1}^{k-1}\frac{1}{2}[\g^{[l]}\stackrel{*}{,}\g^{[k-l]}])\,,
\qquad k\geq 2\,. \label{4.9}
\end{equation}
\end{prop}
\begin{proof}
At order $1$, equation \eqref{3.30} is indeed satisfied since
$\gamma^{[0]} C=0$. Furthermore, the expression
$r^{[2]}=\gamma^{[1]} C+\frac{1}{2}[C\stackrel{*}{,}C]$ depends
only on differentiated ghosts and hence vanishes when the
appropriate variables $y,z$ defined analogously to~\eqref{hatvar}
are zero. Suppose that we have constructed
$\g^{k-1}=\sum_{l=1}^{k-1} \g^{[l]}$ satisfying \eqref{3.30} up to
order $k$, i.e.,
\begin{equation}
\gamma \g^{k-1}+\frac{1}{2}[\g^{k-1}\stackrel{*}{,}\g^{k-1}]
=r^{[k]}+\sum_{m\geq 0} q^{[k+1+m]}\,,
\end{equation}
with $r^{[k]}$ and $\g^{[k]}$, $k\geq 2$ in homogeneity $k$ and
depending only on differentiated ghosts. Applying
$\gamma+[\g^{k-1}\stackrel{*}{,}\cdot]$, the left hand side
vanishes identically, which implies for the lowest order on the
right hand side that $\gamma^{[0]} r^{[k]}=0$.

In homogeneity degree $k$ equation~\eqref{3.30} implies:
\begin{equation}
\gamma^{[0]}\g^{k}=-\gamma^{[1]} \g^{[k-1]}
-\sum_{l=1}^{k-1}\frac{1}{2}[\g^{[l]}\stackrel{*}{,}\g^{[k-l]}]=r^{[k]}\,.
\label{6.24}
\end{equation}
Applying $\rho^{[0]}$ to this equation, it follows from
\eqref{A333} and the induction hypothesis that $r^{[k]}$ does not depend on
undifferentiated ghosts that $\g^{[k]}=-\rho^{[0]} r^{[k]}$ is a
solution to Eq.~\eqref{6.24}. It remains to be checked that
$r^{[k]}$ does not depend on undifferentiated ghosts. Indeed, the
only possible dependence on undifferentiated ghosts can come from
$\gamma^{[1]} \g^{[k-1]} +[C,\g^{[k-1]}]$. This dependence cancels
between the two terms because: (i) $\gamma^{[1]}=
-\commut{C}{V_M}^A \dl{V_M^A}$ up to terms that involve only
differentiated ghosts, with
\begin{equation}
\{V_M^A\}=\{\d_{(\mu_1}\ldots\d_{\mu_{l-1}}A^A_{\mu_{l)}}\,,
\d_{(\mu_1}\ldots\d_{\mu_{l)}}C^A\,,
\d_{(\mu_1}\ldots\d_{\mu_{l-1}}F^{A[0]}_{\mu_{l})\nu} \};
\end{equation}
(ii) by construction $\g^{[k+1]}$ is a polynomial in matrix-valued
fields and their derivatives so that $\commut{C}{\cdot}$ satisfies
Leibnitz rule.
\end{proof}

Explicitly, the quadratic contribution is given by
\begin{multline}
\g^{[2]}=-\rho^{[0]}\Big((\cos\Lambda_{12}-1)C^A
C^B[T_A,T_B]+i\sin\Lambda_{12} C^A C^B\{T_A,T_B\}\Big) =
\\
= -\Big(\frac{\defpar\theta^{\mu\nu}}{2}\frac{\cos\Lambda_{12}-1}
{\Lambda_{12}}A^A_\mu \partial_\nu C^B[T_A,T_B]
+i\frac{\defpar\theta^{\mu\nu}}{2}\frac{\sin\Lambda_{12}}
{\Lambda_{12}} A^A_\mu \partial_\nu
C^B\{T_A,T_B\}\Big)\,.\label{h2}
\end{multline}
\begin{prop}\label{lemma:f-solution-hom}
For $\g=\sum_{k=1}\g^{[k]}$ as in
proposition~\bref{lemma:rec1-hom}, a particular recursive solution
$f_\mu=\sum_{k=1}f_\mu^{[k]}$  of \eqref{3.29} with
$f_\mu^{[1]}=A_\mu$ is given by
\begin{equation}
f_\mu^{[k]}=-\rho^{[0]}\left(\gamma^{[1]}
{f_{\mu}}^{[k-1]}-\partial_\mu
\g^{[k]}-\sum_{l=1}^{k-1}[f_\mu^{[l]}\stackrel{*}{,}\g^{[k-l]}]\right)\,,
\qquad k\geq 2. \label{4.11}
\end{equation}
\end{prop}
\begin{proof} At order $1$, equation \eqref{3.29} is indeed satisfied
since $\gamma^{[0]} A_\mu=\partial_\mu C$. Furthermore,
$-\gamma^{[1]} A_\mu +\partial_\mu \g^{[2]}+
[A_\mu\stackrel{*}{,}C]$ only depends on differentiated ghosts.
Suppose that we have constructed
$f_\mu^{k-1}=\sum^{k-1}_{l=1}f_\mu^{[l]}$ such that~\eqref{3.29}
is satisfied up to order $k$,
\begin{equation}
\gamma f_\mu^{k-1}-\partial_\mu
\g-[f_\mu^{k-1}\stackrel{*}{,}\g]=t_\mu^{[k]}+\sum_{m=0}v_\mu^{[k+1+m]}\,,
\end{equation}
with $t_\mu^{[k]}$ depending only on differentiated ghosts.
Applying $\gamma + \qcommut{\g}{\cdot}$ implies to lowest order
that $\gamma^{[0]} t_\mu^{[k]}=0$. Applying $\rho^{[0]}$, it
follows that $t_\mu^{[k]}=\gamma^{[0]}\rho^{[0]}t_\mu^{[k]}$ which
implies that $f_\mu^{[k]}=-\rho^{[0]} t_\mu^{[k]}$ is a particular
solution to equation~\eqref{3.29} at homogeneity order $k$:
\begin{equation}
  \gamma^{[0]}f_\mu^{[k]}=
-\gamma^{[1]} f_{\mu}^{[k-1]}+\partial_\mu
\g^{[k]}-\sum_{l=1}^{k-1}
[f_{\mu}^{[l]}\stackrel{*}{,}\g^{[k-l]}]=-t_\mu^{[k]}\,.
\end{equation}
It remains to be shown that $t^{[k]}_\mu$ depends only on
differentiated ghosts. Indeed, the only dependence on
undifferentiated ghosts can come from the terms $\gamma^{[1]}
f_{\mu}^{[k-1]}-[f_{\mu}^{[k-1]},C]$ and as before, this
dependence cancels between the two terms. Finally,
$f_\mu^k=f_\mu^{k-1}+f_\mu^{[k]}$ satisfies Eq.~\eqref{3.29} up to
order $k+1$ and the construction can be iterated.
\end{proof}

\subsection{Universal SW map}

The particular SW maps constructed in the previous two sections
depend very little on the associative algebra $\cU$. Indeed, if as
in \eqref{h2}, one does not use the multiplication table in $\cU$
to simplify the expressions, one can construct a "universal" SW
map valid for any $\cU$. A SW map for a particular $\cU$ is then
obtained by using the multiplication table in the final
expression. More precisely, this means that one should first
construct the SW map for Yang-Mills fields taking values in the
free tensor algebra of some sufficiently large vector space.
Because each associative algebra $\cU$ can be represented as a
quotient of such a free tensor algebra modulo some relations, the
SW map for $\cU$ can be obtained by using these relations in the
SW map for the free tensor algebra.

\subsection{Ambiguities in the SW map} Some part of the
arbitrariness in the SW map was discussed originally
in~\cite{Asakawa:1999cu,Bichl:2001cq,Brace:2001rd}. In this
section, we are going to derive the most general solution to the
SW gauge equivalence condition \eqref{3.31}. As before indices in
parentheses refer to the expansion in the deformation parameter.

\begin{prop}\label{lem5}
The general solution $f_\mu^\prime,\g^\prime$ to equations
\eqref{3.29}-\eqref{3.30} with boundary conditions
${f_\mu^\prime}^{(0)}=A_\mu,\,{\g^\prime}^{(0)}=C$ is given by the
composition
\begin{equation}
f_\mu^\prime=f_\mu[f^c[A;\defpar];\defpar]\,,\qquad
\g^\prime=\g[f^c[A;\defpar],\g^c[A,C;\defpar];\defpar]\label{eq:4.1}
\end{equation}
where $f_\mu,h$ is any particular solution with the same boundary
conditions (e.g. the one constructed recursively in the previous
section), while $f^c_\mu,\g^c$ is the general solution to the
"commutative" equations
\begin{align}
  \gamma \g^c+\half\commut{h^c}{h^c}=0\label{eq:comm-1}\,,\\
  \gamma f^c_\mu=\d_\mu \g^c+\commut{f^c_\mu}{\g^c}\label{eq:comm-2}\,,
\end{align}
subject to the boundary condition
\begin{equation}
f_\mu^{c(0)}=A_\mu \,,\qquad \g^{c(0)}=C\,. \label{eq:bound}
\end{equation}
\end{prop}
\begin{proof}
Let $F^\prime=\int dx\,(\hat A^{\mu\star}_A f^{\prime A}_\mu+\hat
C^*_A \g^{\prime A})$ and $F=\int dx\,(\hat A^{\mu\star}_A
f^{A}_\mu+\hat C^*_A \g^A)$ be the generating functionals for the
anticanonical transformation associated with the two SW maps. Denoting
by $F^*$ the  anticanonical transformation corresponding to $F$
(acting on functionals), one has by definition~(see eq.~\eqref{3.35})
\begin{equation}
  F^* (\hat S) = S_0^{\eff}+\sum_{k\geq 1}S^{(0)}_k\,, \qquad
\end{equation}
and similarly for ${F^\prime}^{*}$ with $S_0^{\eff}$ replaced by
some ${S_0^\eff}^\prime$. It then follows that
\begin{equation}
  {F^\prime}^*\Big({F^{-1}}^* \big(\sum_{k\geq 1}S^{(0)}_k\big)\Big)=
  \sum_{k\geq 1}S^{(0)}_k\,+\,\text{antifield-independent terms}\,.
\end{equation}
The antifield dependent part of this equation means that the
anticanonical transformation ${F^\prime}^* \circ {F^{-1}}^*$
preserves the gauge structure of the commutative theory. With
\begin{equation} {F^\prime}^*({F^{-1}}^*(A_\mu))=f^c_\mu[A;\defpar],\qquad
{F^\prime}^*({F^{-1}}^*(C))=\g^c[A;\defpar](C)\,,\end{equation}
the same argument as in~\bref{subsec:def-eq} shows that
equations~\eqref{eq:comm-1} and \eqref{eq:comm-2} hold.
\end{proof}

\begin{prop}\label{prop:comm-sol-gen}
The general solution to equations~\eqref{eq:comm-1}
and~\eqref{eq:comm-2} with boundary conditions~\eqref{eq:bound} is
given by
\begin{align}
  h^c&=(\Lambda^{-1}C\Lambda+\Lambda^{-1}\gamma\Lambda)
  \Big|_{C^A\to C^A+C^Br^A_B,\,\, A^A_\mu\to A^A_\mu
  +W^A_\mu+(A^B_\mu+W^B_\mu)r^A_B}\,,\label{eq:f1}\\
  f^c_\mu&=(\Lambda^{-1}A_\mu \Lambda+\Lambda^{-1}\d_\mu\Lambda)
  \Big|_{A^A_\mu\to A^A_\mu+W^A_\mu+(A^B_\mu+W^B_\mu)r^A_B} \label{eq:f2}\,.
\end{align}
Here $\Lambda=\exp(\lambda^A T_A)$ with $\lambda^A[A;\defpar]$ an
arbitrary formal power series in $\defpar$ with coefficients in
local functions; $r$ has the form $r=C^A r_A^B(\defpar)\, T_B$ and
satisfies
\begin{equation}
\label{eq:MC}
  \bar\gamma r+\half\commut{r}{r}=0\,,
\end{equation}
while $W_\mu=W^A_\mu[A;\defpar]\,T_A$ satisfies
\begin{equation}
  \label{eq:wmu-eq}
  \bar\gamma W_\mu=0\,.
\end{equation}
\end{prop}
The proof is given in appendix {\bf C}. Notice that
equation~\eqref{eq:MC} is the Maurer-Cartan equation for $r$
considered as a $1$-cochain of the Lie algebra $\U$ with
coefficients in the adjoint representation. Notice also that
equation~\eqref{eq:wmu-eq} is equivalent to
$W_\mu=W_\mu(\chi;\defpar)$, where the variables $\chi^A_\Delta$
are defined in appendix {\bf A}, and the following purely
algebraic condition
\begin{equation}
-\commut{C}{\chi^A_\Delta}\,\dd{W_\mu}{\chi^A_\Delta}+\commut{C}{W_\mu}=0\,.
\end{equation}

\begin{theorem}\label{thm:amb}
The general solution $f^\prime_\mu,h^\prime$ to
equations~\eqref{3.29} and \eqref{3.30} with boundary conditions
${f_\mu^\prime}^{(0)}=A_\mu,\,{\g^\prime}^{(0)}=C$ is given by
\begin{equation}
\label{eq:st}
\begin{aligned}
\g^\prime&=\left( \hat\Lambda^{-1}* \g *\hat\Lambda ~+~
\hat\Lambda^{-1}*\gamma \hat\Lambda \right)\big|_{C^A\to
C^A+C^Br^A_B,\,\, A^A_\mu\to A^A_\mu
  +W^A_\mu+(A^B_\mu+W^B_\mu)r^A_B}\,,
\\
f^\prime_\mu&=\left( \hat\Lambda^{-1}*f_\mu * \hat\Lambda~+~
\hat\Lambda^{-1}*\d_\mu \hat\Lambda\right)\big|_{A^A_\mu\to
A^A_\mu +W^A_\mu+(A^B_\mu+W^B_\mu)r^A_B}\,.
\end{aligned}
\end{equation}
Here, $f_\mu,\g$ is a particular solution with the same boundary
conditions; $\hat\Lambda=\exp_*(\hat\lambda^A\,T_A)$ with
$\hat\lambda^A[A,\defpar]$ an arbitrary formal power series in
$\defpar$ with coefficients in local functions; $r=C^A
r_A^B(\defpar)\, T_B$ and $W_\mu=W^A_\mu[A;\defpar]\,T_A$ satisfy
\begin{equation}
\label{eq:nec} \bar\gamma r+\half\commut{r}{r}=0\,,\qquad
\bar\gamma W_\mu=0\,.
\end{equation}
\end{theorem}
The proof is given in appendix {\bf D}. The general SW map
$f^\prime_\mu$ for the gauge fields
can thus be obtained from a particular map $f_\mu$,
by applying a noncommutative gauge transformation, followed by the
substitution $A^A_\mu\to A^A_\mu+A^B_\mu r^A_B$ related
to a Lie algebra automorphism (see below), and then by the
gauge covariant redefinition $A^A_\mu\to A^A_\mu+W^A_\mu$ of
the gauge fields.

The constants $r_A^B$ which enter here define an
automorphism of the Lie algebra (\ref{Liealg}).
This is seen
by rewriting equation~\eqref{eq:MC} in the equivalent form
\begin{equation}
\label{eq:MC'} \gamma \tilde C=-\tilde C\,\tilde C\,,\quad
\tilde C=C+C^A r_A^B T_B\, .
\end{equation}
Using that $\gamma C^A=-\half f_{BC}{}^A C^B C^C$,
one readily verifies that equation~\eqref{eq:MC'} is equivalent to
\begin{equation}
\label{eq:Lauto}
f_{BC}{}^D e_D^A=e_B^D\, e_C^E\, f_{DE}{}^A,\quad
e_A^B=\delta_A^B+r_A^B.
\end{equation}
This establishes indeed a Lie algebra
automorphism $T_A\longrightarrow e_A^B T_B$.
This automorphism is relevant only if it is an
outer automorphism because inner automorphisms can be absorbed by
redefinitions of $\Lambda$. In the $u(N)$ case the only outer
automorphisms that occur here are rescalings of the $U(1)$ generator (since
only automorphisms connected to the identity occur owing
to the boundary condition~\eqref{eq:bound}).

\subsection{Nontriviality of the deformed action. Chern-Simons
theory}\label{s5}

One remaining question is whether the
noncommutative $U(N)$ theory as a whole is equivalent to the
commutative one, i.e., whether the starting point noncommutative
and commutative actions are related through field redefinitions. A
necessary condition for triviality is that, at first order, the
integrand in \eqref{3.22} be on-shell (for the equations of motion
of the commutative theory) equal to a total divergence,
\begin{equation}
 L^{{\rm eff}(1)}_0\approx
\partial_\mu j^\mu_0.\label{trivym}
\end{equation}
This condition can be equivalently rewritten as $L^{{\rm
eff}(1)}_0 d^n x=d j_0+\delta k_1$
 which in turn
implies that
\begin{equation}
L^{{\rm eff}(1)}_0\ d^nx=dJ+s K \label{eq:cons-cond}
\end{equation}
(see e.g. \cite{Barnich:2000zw}). Indeed, it follows from the fact
that $\gamma L^{{\rm eff}(1)}_0=d l_0$ that one can find $K$ and
$\tilde J$ such that $\delta k_1=sK + dJ$. Namely, using the
fact that $\delta f+ d g=0$ implies $f=\delta f_1 + d
g_1$ for any $f$ of positive pure ghost number and antifield number, one
can find $k_2$ such that $\delta k_2+ \gamma k_1=dj_1$. Proceeding further by induction in antifield number one
can construct $K=k_1+k_2+\ldots$ satisfying $\delta k_1=sK +
d J$ for some $J$.

In order to prove non-triviality of the noncommutative deformation
of $U(N)$ Yang-Mills theory, it is enough to consider the $U(1)$
case. Indeed, by putting to zero all the components of the gauge
field except for the one associated to the $u(1)$ factor,
triviality of the $U(N)$ theory would imply triviality of the
$U(1)$ theory. By applying $s$ to the
equation~\eqref{eq:cons-cond}, we get the descent equation
$sJ+d m=0$ for the $n-1$ form $J$. This last equation then
implies, in space-time dimensions strictly greater than $2$, (see
eq.~(13.6) of \cite{Barnich:2000zw}) that
\begin{equation}
\label{eq:j-general} J=\lambda(\star A^* C+\star  F A) + A
P(F)+I^{n-1}(\chi)+ s (\cdot)+d(\cdot)\,,
\end{equation}
where $F=F_{\mu\nu}dx^\mu dx^\nu, A=A_\mu dx^\mu$, $\star$ denotes
hodge conjugation, and $P(F)$ is a polynomial in the two-form $F$.

It then follows from \eqref{eq:j-general} that
\begin{equation}
L^{{\rm eff}(1)}_0\ d^nx \approx \lambda (-)^n \star F F+F P(F)+d
I^{n-1}(\chi)\,.
\end{equation}
Since $L^{{\rm eff}(1)}_0$ is cubic in $A_\mu$ only terms of
homogeneity degree $3$ can contribute:
\begin{equation}
L^{{\rm eff}(1)}_0\ d^nx \approx \alpha F^3+d I^{n-1}(\chi).
\end{equation}
In the case where there is no explicit $x$ dependence, this
equation implies, by putting to zero the derivatives of the field
strength, that
\begin{equation}
L^{{\rm eff}(1)}_0\ d^nx=\alpha F^3.
\end{equation}
Because this is not the case, we conclude that the noncommutative
deformation of the action is non trivial, at least in the $x$
independent case and in space-time dimension strictly greater than
$2$.

This is to be contrasted with noncommutative $U(N)$ Chern-Simons
theory described by the action
\begin{equation}
 \hat S^{CS}_0=\int\ {\rm Tr}\ (\hat A*d\hat A +\frac{2}{3}\hat A*\hat
A*\hat A).
\end{equation}
The antifield structure of the solution of the master equation is
the same as in the Yang-Mills case. In \cite{Grandi:2000av}, it
has been shown that the SW map transforms the noncommutative
$U(N)$ Chern-Simons theory into its commutative counterpart, so
that the noncommutative deformation is trivial.

{}From the point of view of the local BRST cohomology of the
commutative theory, this follows directly from the following
arguments: for Chern-Simons theory with gauge group $U(N)$, there exists exactly one
cohomology class in form degree $n$ and ghost number $0$ in the
case where $N\geq 2$, and none in the $U(1)$ case (see e.g.
\cite{Barnich:2000zw}.) If we denote by $g_{CS}$ the gauge
coupling constant that goes with the structure constants of the
$su(N)$ subalgebra, this representative can be chosen to be
$\frac{\partial S^{(0)}}{\partial g_{CS}}$. It follows that any
consistent deformation of commutative $U(N)$ Chern-Simons theory
can be absorbed by a field-antifield redefinition and a
redefinition of the gauge coupling. In particular, it follows that
there exists a field-antifield redefinition such that \bea \hat
S^{CS}[\hat \phi[\phi,\phi^*;\defpar,g_{CS}],\hat \phi^*
[\phi,\phi^*;\defpar,g_{CS}];\defpar,\hat g_{CS}]= S^{(0) CS
}[\phi,\phi^*; g_{CS}(\defpar)], \eea with
$g_{CS}(\defpar)=g_{CS}+\defpar f(g_{CS})+\dots$. In the case
where there is no explicit $x$ dependence, dimensional arguments
imply that $g_{CS}(\defpar)=g_{CS}$.

\section{Local BRST cohomology of noncommutative $U(N)$ Yang-Mills theory}\label{sec:coh}

In this section, we discuss the local BRST cohomology groups of
noncommutative $U(N)$ Yang-Mills theory since these groups contain
information about anomalies, counterterms, observables, symmetries
and conservation laws of the theory (for details see
\cite{Barnich:2000zw} and references therein). By using a SW map,
these groups can be most conveniently analyzed in terms of the
effective Yang-Mills theory because only the part of the BRST
cohomology groups that pertains to the dynamics differs from that
of standard Yang-Mills theory. 
The BRST cohomology
groups for effective Yang-Mills theories have been
discussed 
in \cite{Barnich:2000zw}.

\subsection{Basic considerations}

Let $\Omega$ be the space of local forms and
$\Omega[[\defpar]]=\Omega\otimes [[\defpar]]$ be the space of
formal power series in the deformation parameter with values in
$\Omega$.  We want to analyze $H^{k,p}(\hat
s|d,\Omega[[\defpar]])$, the local BRST cohomology group in ghost
number $k$ and form degree $p$. An element of $H^{k,p}(\hat
s|d,\Omega[[\defpar]])$ is an equivalence class of a local
$p$-form $a^{k,p}$ of ghost number $k$ satisfying the cocycle
condition
\begin{equation}
\label{eq:basic-cons}
  \hat s a^{p,k} +d a^{p+1,k-1}=0\,,
\end{equation}
modulo the equivalence relation
\begin{equation}
\label{eq:basic-rel} a^{k,p} \sim a^{k,p}+\hat s b^{k-1,p} +d
b^{k,p-1}\,.
\end{equation}

Standard arguments using an expansion in the deformation parameter
allow one to show that $H^{k,p}(\hat s|d,\Omega[[\defpar]])\subset
H^{k,p}(\hat s^{(0)}|d,\Omega)\otimes[[\defpar]]$. In order to
describe $H^{k,p}(\hat s|d)$ it is thus sufficient to check which
elements from $H^{k,p}(\hat s^{(0)}|d,\Omega)$ can be completed to
elements in $H^{k,p}(\hat s|d,\Omega[[\defpar]])$. The same
considerations apply to the groups $H(\hat
\gamma|d,\Omega[[\defpar]])$ and $H(\hat
\delta|d,\Omega[[\defpar]])$.

Let $\hat z^A$ denote collectively all the fields, ghosts, and
antifields of the theory. Consider the invertible change of
variables $\hat z^A=\hat z^A[z;\defpar]$, with $\hat
z^A[z;\defpar]$ formal power series in $\defpar$ with values in
local functions depending on the new variables $z^B$. This change
of variables is extended to the derivatives $\hat
z^A_{,\mu_1\dots\mu_k}$ in such a way that the expression of the
total derivative $\partial_\mu=\dd{}{x^\mu}+\hat
z^A_{,\mu}\dd{}{\hat z^A}+\dots$ is the same in the new variables
$z^B$ as it was in the old variables. In terms of the new
variables, the expression for the differential $\hat s$ becomes
\bea \hat s z^B=\sum_{k=0}\Big[(\d_{\mu_1}\dots \d_{\mu_k} \hat s
\hat z^A)(\dd{z^B}{\hat z^A_{,\,\mu_1\dots\mu_k}})\Big]\Big|_{\hat
z=\hat z[z]}.\label{changevar} \eea

\subsection{Local BRST cohomology in the effective theory}

Let us now apply the considerations of the previous subsection to
the case of noncommutative $U(N)$ Yang-Mills theory and use as a
change of variables a particular SW map. We denote the BRST
differential in terms of the new unhatted variables by $s^{\rm
eff}$. By definition of a SW map, $s^{\rm eff}=\delta^{\rm
eff}+\gamma$. The differentials $\gamma$ and $\delta^{\rm eff}$
coincide with $\hat \gamma^{(0)}$ respectively $\hat \delta^{(0)}$
in terms of unhatted fields and antifields, except for the action
of $\delta^{\rm eff}$ on the antifields $A^{*\mu}_A$, which
contains correction terms due to the higher dimensional gauge
invariant interactions, \bea \delta^{\rm eff}
A^{*\mu}_A=\frac{\delta S^{\rm eff}_0[A]}{\delta
A^A_\mu}=\frac{\delta S^{(0)}_0[A]}{\delta
A^A_\mu}+O(\defpar).\eea

As a consequence, local BRST cohomology groups like
$H(\hat\gamma|\,d,\Omega[[g]])$ that do not involve the dynamics,
are completely determined by the local BRST cohomology groups of
standard $U(N)$ Yang-Mills theory,
\begin{equation}
H(\hat\gamma|d,\Omega[[g]]) \simeq H(\gamma|d,\Omega)\tensor
[[\defpar]]\,.
\end{equation}
In particular, in ghost number $1$ and form degree $n$, this means
that the potential chiral anomalies of noncommutative $U(N)$
Yang-Mills theory (treated as an effective theory) are directly
related to the commutative ones.

How the dynamics of effective Yang-Mills theories enters the local
BRST cohomology groups has been analyzed in some detail in
\cite{Barnich:2000zw}. In particular, this analysis also applies
to the effective field theory formulation of noncommutative $U(N)$
Yang-Mills theories.

In the following subsection, we will only briefly discuss some
BRST cohomology classes of the standard $U(N)$ Yang-Mills theory
that involve the dynamics and get obstructed under the
noncommutative deformation.

\subsection{Breaking of Poincar\'e invariance in
noncommutative deformation}

Local BRST cohomology classes that involve the dynamics are for
instance those in ghost number $-1$ and form degree $n$. They
contain the information about the global symmetries and the
associated conserved currents. In particular, we concentrate on
the Poincar\'e invariance, or, in 4 space-time dimensions, the
conformal invariance of standard Yang-Mills theory.

Standard homological arguments (see e.g.~\cite{Barnich:2000zw})
show that
\begin{equation} H^{-1,n}(\hat s|d,\Omega[[\defpar]])\simeq
H^n_1(\hat\delta|d,\Omega[[\defpar]])\,.\end{equation} In turn, by
the same reasoning that relates the $\hat s$ to the $s^{\rm eff}$
cohomology, this group is given by
$H^n_1(\hat\delta|d,\Omega[[\defpar]])\simeq H^n_1(\delta^{\rm
eff} |d,\Omega[[\defpar]])$. For representatives
$\omega_1^n=Q^A_\mu A^{*\mu}_Ad^nx$, the cocycle condition of this
last group determines the global symmetries and the associated
Noether currents:
\begin{equation}
 \delta^{\rm eff}(A^{*\mu}_AQ^A_\mu d^nx)+d j=0\
 \iff \ \frac{\delta L^{\rm eff}}{\delta A^A_{\mu}}Q^A_{\mu}+\d_\mu
 j^\mu=0\,.\label{6.6}
\end{equation}
By taking the coboundary condition into account,
$H^n_1(\delta^{\rm eff} |d,\Omega[[\defpar]])$ can be shown to
correspond to the non trivial global symmetries (respectively the
non trivial conserved currents) associated to $L^{\rm eff}_0$.

In space-time dimensions $n$ strictly greater than $2$, the
general solution to the conformal Killing equation
\begin{equation}
\label{eq:conf-kil-cond} \eta_{\mu\rho}\d_\nu
\xi^\rho+\eta_{\nu\rho}\d_\mu\xi^\rho=\frac{2}{n}\eta_{\mu\nu}\d_\rho
\xi^\rho\,,
\end{equation}
reads as
\begin{equation}
  \xi^{\mu}=a^\mu+\lambda^\mu_\nu x^\nu + c x^\mu +
  b^\nu \eta_{\nu\rho}x^\rho x^\mu - \half b^\mu \eta_{\nu\rho} x^\nu
  x^\rho\,.
\end{equation}
Here, the constants $a^\mu$ correspond to translations,
$\lambda^\mu_\nu$ with $\lambda_{\mu\nu}\equiv
\eta_{\mu\rho}\lambda^\rho_\nu=-\lambda_{\nu\mu}$ to Lorentz
transformations, $c$ to dilatations and $b^\mu$ to special
conformal transformations. The conformal transformations act on
the potentials and the associated curvatures as Lie derivatives,
\begin{equation}
  \begin{split}
  \delta_\xi A_\mu &= \lied_{\xi} A_{\mu}=\xi^\rho \d_\rho A_\mu + A_\rho \d_\mu \xi^\rho\,,\\
  \delta_\xi F_{\mu\nu} &= \lied_{\xi} F_{\mu\nu}=\xi^\rho \d_\rho F_{\mu\nu}+F_{\rho \nu}\d_\mu
  \xi^{\rho}+ F_{\mu\rho}\d_\nu \xi^{\rho}\,.
\end{split}
\end{equation}
Because the Poincar\'e, respectively the conformal transformations
in 4 dimensions, are symmetries of commutative Yang-Mills theory,
taking $Q^{(0)A}_\mu=\delta_\xi A^A_\mu$ with the appropriate
$\xi$ allows one to satisfy equation \eqref{6.6} to order zero in
$\defpar$ for some $j^{(0)\mu}$.

In order for these symmetries to survive the noncommutative
deformation to first order in $\defpar$, one needs to find
$Q^{(1)A}_\mu, j^{(1)\mu}$ such that
\begin{equation}
\label{eq:1st-cons} \vddl{L_0^{\rm eff (1)}}{A^A_\mu}\delta_\xi
A_{\mu}+\vddl{L_0^{\rm eff (0)}}{A^A_\mu}Q^{(1)A}_\mu+\d_\mu
j^{(1)\mu}=0.
\end{equation}
Explicitly, the first term reduces to
\begin{multline}
\vddl{L_0^{\rm eff (1)}}{A^A_\mu}\delta_\xi A_{\mu} = \frac{i
\eta^{\mu\rho}\eta^{\nu\sigma}}{2\kappa^2} \left(
\theta^{\alpha\sigma}\d_\sigma
\xi^\beta+\theta^{\sigma\beta}\d_\sigma \xi^\alpha \right)
\\
\tr \left( -\half
F_{\mu\nu}\cscommut{F_{\alpha\rho}}{F_{\beta\sigma}}+ \frac{1}{8}
F_{\alpha\beta} \cscommut{F_{\mu\nu}}{F_{\rho\sigma}} \right)\,,
\end{multline}
where we used that $\xi$ is a conformal Killing vector. This
expression coincides with $L_0^{\rm eff(1)}$ with
$\theta^{\alpha\beta}$ replaced by $\theta^{\alpha\sigma}\d_\sigma
\xi^\beta+\theta^{\sigma\beta}\d_\sigma \xi^\alpha$. In its turn,
$L_0^{\rm eff(1)}$ was proved in subsection \bref{s5} to be non
trivial, i.e., not proportional to equations of motion modulo a
total derivative\footnote{For the special conformal
transformations, the arguments given in subsection \bref{s5} have
to be extended to local forms depending explicitly on $x$ since
$\6_\mu \xi^\nu$ depends linearly on $x$.}. Thus we conclude that
Eq.~\eqref{eq:1st-cons} admits solutions $Q^{(1)A}_\mu,
j^{(1)\mu}$ iff
\begin{equation}
\label{eq:theta-pres} \theta^{\alpha\sigma}\d_\sigma
\xi^\beta+\theta^{\sigma\beta}\d_\sigma \xi^\alpha=0\,,
\end{equation}
i.e., if $\xi$ preserves $\theta$.

Let us assume that $\theta$ is nondegenerate (symplectic). It then
follows that $\d_\mu\xi^\mu=0$. This means in particular in 4
dimensions that the dilatations and the special conformal
symmetries are obstructed. Thus we can assume that
\begin{equation}
\xi^{\mu}=a^\mu+\lambda^\mu_\nu x^\nu\,.\label{eq:NC-Kiling}
\end{equation}
The Killing condition together with \eqref{eq:theta-pres} then
require
\begin{equation}
\label{eq:eta-theta-pres}
  \begin{split}
        \eta^{\mu\rho}\lambda^\nu_\rho+\eta^{\rho\nu}\lambda^\mu_\rho&=0\,,\\
        \theta^{\mu\rho}\lambda^\nu_\rho+\theta^{\rho\nu}\lambda^\mu_\rho&=0\,.
\end{split}
\end{equation}
Conversely, any $\xi$ of the form $\xi^\mu=a^\mu+\lambda^\mu_\nu
x^\nu$ satisfying the previous equations defines a symmetry of
$L^{\rm eff}_0$. Indeed, since $\xi$ is at most linear in $x$, we
have $\commut{L_{\xi}}{\d_\mu}=0$. For any function
$f(\eta,\theta,[A])$ with all space-time indices of $A^A_\mu$ and
their derivatives contracted by either $\theta^{\mu\nu}$ or
$\eta^{\mu\nu}$, we then get \begin{equation}\delta_\xi
f=\sum_{k=0}\d_{\rho_1}\dots\d_{\rho_k}(\lied_\xi
A^A_\mu)\frac{\partial f}{\partial A^A_{\mu,\rho_1\dots\rho_k}}
=\lied_\xi f=\d_\rho(\xi^\rho f)\,,\label{6.15}\end{equation}
where the second equality holds because
$\lied_\xi\theta=0=\lied_\xi\eta$ on account of
\eqref{eq:eta-theta-pres} and the last equality follows from
$\d_\rho\xi^\rho = 0$. Thus $L^{\rm eff}_0$ is invariant up to a
total derivative to all orders in $\defpar$, which means that
\eqref{6.6} can be satisfied with $Q^A_\mu=Q^{(0)A}_\mu=\delta_\xi
A^A_\mu$. Hence, only the Poincar\'e transformations that satisfy
in addition $\lied_\xi\theta=0$ are unobstructed and define global
symmetries of $L^{\rm eff}_0$.

In fact, the argument leading to \eqref{6.15} can be used to show
directly without using the SW map that, in the case where $\theta$
is non degenerate, the Poincar\'e transformations satisfying
$\lied_\xi\theta=0$ define global symmetries of $\hat L_0$.

Let us now discuss in more detail the Lie subalgebra of Lorentz
transformations that satisfy $\lied_\xi\theta=0$ for a non
degenerate $\theta$. This Lie subalgebra crucially depends on the
respective position of the matrices $\eta^{\mu\nu}$ and
$\theta^{\mu\nu}$. It is useful to introduce the linear operators
\begin{equation}
  J_\mu^\nu=\theta_{\mu\rho}\eta^{\rho\nu}\,,\qquad
  K_\mu^\nu=J_\mu^\rho J_\rho^\nu\,,\label{616}
\end{equation}
with $\theta_{\mu\rho}\theta^{\rho\nu}=\delta^\nu_\mu$ and
$K_{\mu\nu}\equiv \eta_{\mu\rho}K^\rho_\nu=K_{\nu\mu}$.
\begin{prop}\label{prop:nondeg-symm}
If the operator $K$ can be diagonalized over ${\mathbb R}$, with
$t_1,\dots, t_N$ denoting its distinct eigenvalues, then the
eigenspaces of $K$ with eigenvalues $t_\alpha$ are
even-dimensional of dimension $2n_\alpha$ and the subalgebra
$\mathfrak g$ of Lorentz transformations satisfying
$\lied_\xi\theta=0$ decomposes as  \bea {\mathfrak g}\simeq
\oplus_{\alpha=1}^N {\mathfrak g}_\alpha,\eea where for negative
$t_\alpha$, ${\mathfrak g}_\alpha$ is $u(n^+_\alpha,n^-_\alpha)$
with $n^+_\alpha+n^-_\alpha=n_\alpha$, while for positive
$t_\alpha$, ${\mathfrak g}_\alpha$ is $gl(n_\alpha,{\mathbb R})$.
\end{prop}
The proof of the proposition is given in the appendix~{\bf E}. In
particular, if $\eta$ is Euclidean, $K$ can be diagonalized and
\begin{equation}
{\mathfrak g}\simeq\oplus_{\alpha=1}^N u(n_\alpha)\,, \qquad
\sum_{\alpha=1}^N n_\alpha=\frac{n}{2},
\end{equation}
where $n$ denotes the dimension of the Euclidean space (which is
even because $\theta$ is non degenerate). If all the eigenvalues
of $K$ coincide, the symmetry algebra is $u(n/2)$ of maximal
dimension $n^2/4$.

In the 4-dimensional Minkowski case with canonical $\theta$,
\begin{equation}
\eta = \left(
  \begin{array}{cccc}
 -1&0&0&0\\
  0&1&0&0\\
  0&0&1&0\\
  0&0&0&1
  \end{array}
\right)\,, \qquad \theta = \left(
  \begin{array}{cccc}
  0&0&1&0\\
  0&0&0&1\\
 -1&0&0&0\\
  0&-1&0&0
  \end{array}
\right)\,, \qquad
\end{equation}
$K$ is diagonalizable and ${\mathfrak g}\simeq gl(1,{\mathbb R})
\oplus u(1)$ is a 2-dimensional Abelian algebra.

To complete the discussion let us briefly consider the case where
$\theta$ is degenerate. This case is more difficult because there
is no longer a suitable tensor $J$ as in \eqref{616}. For
simplicity, let us discuss the case where the matrices
$\eta^{\mu\nu}$ and $\theta^{\mu\nu}$ take the following form:
\begin{equation}
\eta^{\mu\nu} = \left(
  \begin{array}{cccc}
   \eta^{ij}&0\\
    0&\eta^{ab}
  \end{array}
\right)\,, \qquad \theta^{\mu\nu} = \left(
  \begin{array}{cccc}
  0&0\\
  0&\theta^{ab}\\
  \end{array}
\right)\,.
\end{equation}
This covers in particular Minkowski space-time where the time
coordinate is ``commuting", i.e., for $\eta={\rm diag}
(-1,1,\dots,1)$ and $\theta^{0\mu}=0$. The
condition~\eqref{eq:theta-pres} takes the form
\begin{equation}
   \theta^{ac}\d_c \xi^b+\theta^{cb}\d_c \xi^a=0\,, \qquad \d_c \xi^i=0\,.
\end{equation}
The first equation can be solved by $\xi^a=\theta^{ac}\d_c H$ for
some function $H$.  At the same time conformal Killing
condition~\eqref{eq:conf-kil-cond} implies that
\begin{equation}
(\d_i \xi^i+\d_c \xi^c)\eta^{ab}=\frac{2}{n}(\eta^{ac}\d_c
\xi^b+\eta^{cb}\d_c \xi^a)\,, \qquad  \d_i\xi^c=0\,.
\end{equation}
Multiplying the first equation by $\eta_{ab}$ and substituting
$\xi^a=\theta^{ac}\d_c H$ one gets
\begin{equation}
\d_\mu \xi^\mu=\d_i \xi^i+\d_c \xi^c=0\,,
\end{equation}
which again implies that dilatations and special conformal
symmetries are obstructed in 4 dimensions. The problem can now be
reduced to the non-degenerate case already solved in
proposition~\bref{prop:nondeg-symm}: the algebra $\lieg$ is the
direct sum $\lieg=\lieg_1\oplus \lieg_2$, where $\lieg_1$ are the
Lorentz transformations associated to the metric $\eta^{ij}$ and
$\lieg_2$ is the subalgebra of Lorentz transformations preserving
both $\eta^{ab}$ and the non-degenerate $\theta^{ab}$.

\section*{Acknowledgments}
\addcontentsline{toc}{section}{Acknowledgments} The work of  GB
and MG is supported in part by the ``Actions de Recherche
Concert{\'e}es'' of the ``Direction de la Recherche
Scientifique-Communaut{\'e} Fran\c{c}aise de Belgique", by a
``P{\^o}le d'Attraction Interuniversitaire'' (Belgium), by
IISN-Belgium, convention 4.4505.86, by the INTAS grant 00-00262,
and by the European Commission RTN program HPRN-CT00131, in which
the authors are associated to K.~U.~Leuven. GB is also supported
by Proyectos FONDECYT 1970151 and 7960001 (Chile), while MG is
supported by RFBR grants 01-01-00906 and 02-01-06096.

\section*{Appendices}
\addcontentsline{toc}{section}{Appendices}

\subsection*{Appendix A: Proof of \eqref{eq:S-represent}} \label{a1}

\addcontentsline{toc}{subsection}{Appendix A: Proof of
\eqref{eq:S-represent}}

\def\theequation{A.\arabic{equation}}
\setcounter{equation}{0}

Using explicitly the decomposition according to the antifield
number, defined by assigning degree 1 and 2 to $A^{*\mu}_A$ and
$C^*_A$ respectively, with $A^A_\mu$ and $C^A$ carrying zero
degree, the commutative BRST differential $s$ decomposes as
\begin{equation}
 s=\delta+\gamma,
\end{equation}
where $\delta$ and $\gamma$ are of antifield numbers $-1$ and 0,
respectively. Explicitly,
\begin{alignat}{2}
\label{eq:s}
     \gamma  A_\mu &=  D_\mu  C\,,
 \qquad&
     \gamma  C &= -  C C\,, \\
     \gamma  A^{*\mu}_A &=\f BAC C^B A^{*\mu}_C\,,
 \qquad&
     \gamma  C^*_A &= \f BAC  C^B C^*_C\,,\\
     \delta  A_\mu &= 0\,,
  \qquad&
     \delta  C &= 0\,, \\
     \delta  A^{*\mu}_A &=\displaystyle{\frac{1}{\kappa^2}}
    {\rm Tr}\,(T_A  D_\nu  F^{\nu\mu})\,,
  \qquad&
     \delta  C^*_A &= - \6_\mu  A^{* \mu}_A
     +\f BAC A^B_\mu A^{* \mu}_C\,,
\end{alignat}
where $\f ABC$ are the structure constants defined by $
{}[T_A,T_B]=\f ABC T_C\, .$

An element $\tilde S^{(k+1)}\in \mathfrak S$ contains parts with
antifield numbers 0, 1 and 2: \bea \tilde S^{(k+1)}&=&\tilde
S^{(k+1)}_0+\tilde S^{(k+1)}_1+\tilde S^{(k+1)}_2\,.
\label{ord1d1}\eea Eq.\ (\ref{coc0}) decomposes thus into terms
with antifield numbers 2, 1 and 0: \begin{equation} \gamma\,\tilde
S^{(k+1)}_2=0,\ \gamma\,\tilde S^{(k+1)}_1+\delta\,\tilde
S^{(k+1)}_2=0,\ \gamma\,\tilde S^{(k+1)}_0+\delta\,\tilde
S^{(k+1)}_1=0. \label{coc21}
\end{equation}
Along the lines of \cite{Dubois-Violette:1992ye,Torre:1995kb}, we
introduce the following variables $y^\alpha,z^\alpha,w^i$ as new
coordinates in the space of fields, antifields and their
derivatives:
\begin{equation}
  \begin{aligned}
\{y^\alpha\}&=\{\partial_{(\nu_1}\dots\partial_{\nu_l}
A^A_{\mu)}\}\,,\quad \{z^\alpha\}=
\{\partial_{(\nu_1}\dots\partial_{\nu_l}  D_{\mu)}  C^A\},
\\
\{w^i\}&= \{C^A, D_{(\nu_1}\dots D_{\nu_{l}}
F^{{}A}_{\mu)\lambda}, D_{(\nu_1}\dots  D_{\nu_{l})} A^{*\mu}_A\,,
 D_{(\nu_1}\dots  D_{\nu_{l})} C^*_A\},
\label{var}
  \end{aligned}
\end{equation}
where $l=0,1,\dots$. These variables are independent and complete
in the sense that every local function of the fields, antifields
and their derivatives can be uniquely expressed
in terms of them.%
\footnote{It is the independence of these variables which avoids
constraints as encountered in \cite{Brace:2001fj,Brace:2001rd}. }
We define a homotopy operator $\rho$ on functions of these
variables by \bea
  \rho f(y,z,w)=\int^1_0\frac{dt}{t}\ y^{\alpha}\
  \frac{\partial f(ty,tz,w)}{\partial z^\alpha}\ .
\label{rho} \eea It satisfies
\begin{equation}
\{\gamma,\rho\} f(y,z,w)= f(y,z,w)-f(0,0,w)\,. \label{contract1}
\end{equation}
When $f(y,z,w)$ is $\gamma$-closed this relation yields in
particular: \bea \gamma f(y,z,w)=0\quad\Rightarrow\quad
f(y,z,w)=f(0,0,w)+\gamma\rho f(y,z,w). \label{contract21} \eea It
states that the cohomology of $\gamma$ can be constructed solely in
terms of the $w$'s -- the dependence of $\gamma$-cocycles on the
$y$'s and $z$'s is trivial. We stress that this holds for local
functions -- in general it does not hold for local functionals
whose integrands are $\gamma$-closed only up to total derivatives,
the reason being that $\rho$ does not commute with $\6_\mu$.
Nevertheless we can use (\ref{contract21}) to analyze Eqs.\
(\ref{coc21}) thanks to the fact that $\tilde S^{(k+1)}_2=\int d^n
x\, \omega_2^{(k+1)}$ and $\tilde S^{(k+1)}_1$ are linear in
antifields. Indeed, consider the first equation of (\ref{coc21}).
Since the integrand $\omega_2^{(k+1)}$ is linear in the
undifferentiated antifields $C^*_A$, so is $\gamma \omega_2^{(k+1)}$
(owing to $\gamma C^*_A=\f BAC C^BC^*_C$). Hence  $\gamma
\omega_2^{(k+1)}$ cannot be a (nonvanishing) total derivative and
$\gamma \tilde S^{(k+1)}_2=0$ implies thus that it vanishes, $\gamma
\omega^{(k+1)}_2=0$. Because $\omega^{(k+1)}_2$ depends on the
ghosts only via their derivatives, it vanishes at
$y^\alpha=z^\alpha=0$ when expressed in terms of the variables
$y^\alpha,z^\alpha,w^i$ because $\d_\alpha C=D_\alpha
C-[A_\alpha,C]$ itself vanishes at $y^\alpha=z^\alpha=0$. Using
(\ref{contract21}) we conclude
$\omega^{(k+1)}_2=\gamma\rho\,\omega^{(k+1)}_2$, which yields \bea
\tilde S^{(k+1)}_2=\gamma\tilde\Xi^{(k+1)}_2,\quad
\tilde\Xi^{(k+1)}_2=\int d^nx\,\rho\,\omega^{(k+1)}_2\in \mathfrak
S . \label{coc41} \eea
Using this in the second equation
(\ref{coc21}), the latter yields \bea \gamma(\tilde
S^{(k+1)}_1-\delta\tilde\Xi^{(k+1)}_2)=0\,, \label{coc5} \eea owing
to $\{\delta,\gamma\}=0$. The functional $\tilde
S^{(k+1)}_1-\delta\tilde\Xi^{(k+1)}_2=\int d^nx\ \tilde\eta^{(k+1)}_1$
depends linearly on the antifields $A^{* \mu}_A$ and their
derivatives (derivatives of $A^{* \mu}_A$ occur because
$\delta\Xi^{(1)}_2$ contains $\delta C^*_A=-\6_\mu  A^{* \mu}_A+\dots$).
Using integration by parts
that remove all derivatives of $A^{*
\mu}_A$, this functional can be seen to belong to $\mathfrak S$
and takes the form $\int d^nx\ \eta^{(k+1)}_1$ with
$\eta^{(k+1)}_1$ not involving any derivatives of
$A^{*\mu}_A$.

As before, from $\gamma\int d^nx\,\eta^{(k+1)}_1=0$ we conclude
that $\gamma\eta^{(k+1)}_1=0$ because $\eta^{(k+1)}_1=0$ is linear
in antifields and does not contain any derivatives of them. One
then deduces that $\eta^{(k+1)}_1$ vanishes at
$y^\alpha=z^\alpha=0$ when expressed in terms of the variables
$y^\alpha,z^\alpha,w^i$ because it depends on the ghosts only via
their derivatives. Hence we can use (\ref{contract21}) again, and
conclude that $\eta^{(k+1)}_1=\gamma\rho\,\eta^{(k+1)}_1$ which
yields \bea \tilde S^{(k+1)}_1-\delta\tilde\Xi^{(k+1)}_2 =
\gamma\tilde\Xi^{(k+1)}_1,\qquad\tilde\Xi^{(k+1)}_1&=& \int d^nx\
\rho\,\eta^{(k+1)}_1\in \mathfrak S. \label{additi}\eea Using
eqs.\ (\ref{coc41}) and (\ref{additi}) in (\ref{ord1d1}), we
obtain
\begin{equation}
 \tilde S^{(k+1)}=S_0^{{\rm eff} (k+1)}[A] +
s\tilde\Xi^{(k+1)}\,, \label{coc9}
\end{equation}
with
\begin{equation}
\begin{aligned}
\tilde\Xi^{(k+1)}&=&\tilde\Xi^{(k+1)}_1+\tilde\Xi^{(k+1)}_2\in \mathfrak
S\,,
\\
S_0^{{\rm eff} (k+1)}[A]&=&\tilde S^{(k+1)}_0-
\delta\tilde\Xi^{(k+1)}_1\,.
\end{aligned}
\end{equation}

\subsection*{Appendix B: Explicit construction of $\hat\Xi$ and
$\hat B_0$}
\def\theequation{B.\arabic{equation}}
\setcounter{equation}{0}

\addcontentsline{toc}{subsection}{Appendix B: Explicit
construction of $\hat\Xi$ and $\hat B_0$}

The decomposition of ${\6\hat S}/{\6\defpar}$ according to the
antifield number reads: \bea \dd{\hat S}{\defpar}&=& \dd{\hat
S_0}{\defpar} +\dd{\hat S_1}{\defpar} +\dd{\hat S_2}{\defpar}\,,
\label{ordd}\\
\dd{\hat S_0}{\defpar}&=& \frac{-i\theta^{\alpha\beta}}{2\kappa^2}
\int d^nx\ {\rm Tr}\, (\7F^{\mu\nu}\,\d_\alpha
\7A_\mu\STAR\d_\beta \7A_\nu),
\label{ordc}\\
\dd{\hat S_1}{\defpar}&=& \frac{i\theta^{\alpha\beta}}{2} \int
d^nx\ \7A^{* \mu}_A\,\scommut{\d_\alpha \7A_\mu}{\d_\beta \7C}^A,
\label{ordb}\\
\dd{\hat S_2}{\defpar}&=& \frac{i\theta^{\alpha\beta}}{2} \int
d^nx\ \7C^{*}_A\, (\d_\alpha \7C\STAR\d_\beta \7C)^A, \label{orda}
\eea while $\hat s \dd{\hat S}{\defpar}=0$ decomposes according
to: \bea \7\gamma\ \dd{\hat S_2}{\defpar}=0,\quad \7\gamma\
\dd{\hat S_1}{\defpar}+\7\delta\ \dd{\hat S_2}{\defpar}=0,\quad
\7\gamma\ \dd{\hat S_0}{\defpar}+\7\delta\ \dd{\hat
S_1}{\defpar}=0. \label{full2} \eea In order to analyze these
equations, we use a decomposition according to the homogeneity in
all (hatted) fields and antifields.

We start from the first equation of \eqref{full2}. To remove the
integral, we take the variational derivative with respect to
$\7C^*_A$. This yields \bea 0=\frac{\delta}{\delta
\7C^*_A}\Big[\7\gamma\, \dd{\hat S_2}{\defpar} \Big]=
(\7\gamma\eta+\qcommut{\7C}{\eta})^A, \label{B6} \eea where \bea
\eta= \frac{i\theta^{\alpha\beta}}{2}\, \d_\alpha \7C\STAR\d_\beta
\7C. \label{B5} \eea At lowest order in the fields, Eq.\
(\ref{B6}) yields: \bea \hgz\eta=0. \label{B6a} \eea For $\eta$ of
the form \eqref{B5}, we will show below that $\hgz$ can be
``inverted'', i.e. that there exists $\xi$ such that \bea \eta=
\hgz\xi.\label{B'1} \eea This equation defines $\xi$ only up to a
$\hgz$-cocycle. A convenient choice, to be discussed in more
detail below, turns out to be  \bea \xi=
\frac{i\theta^{\alpha\beta}}{4}\,
\scommut{\7A_\alpha}{\6_\beta\7C}. \label{B7} \eea That this $\xi$
actually satisfies \eqref{B'1} can be directly checked. We thus
obtain \begin{equation}{\6\hat S_2}/{\6\defpar}=\int d^nx\,
\7C^*_A\, \eta^A =\hgz \int d^nx\, \7C^*_A\,
\xi^A\,,\end{equation} where we can replace $\hgz$ with $\7\gamma$
because of $\hgo\int d^nx\, \7C^*_A\,\xi^A=0$ (the latter holds
because $\hgo\xi=-\qcommut{\7C}{\xi}$): \bea \dd{\hat
S_2}{\defpar}=\7\gamma\, \7\Xi_2,\quad \7\Xi_2=
\frac{i\theta^{\alpha\beta}}{4}\int d^nx\  \7C^*_A\,
\scommut{\7A_\alpha}{\6_\beta\7C}^A. \label{B8} \eea Using
(\ref{B8}) in the second equation (\ref{full2}), we obtain: \bea
\7\gamma\,\Big[ \dd{\hat S_1}{\defpar}-\7\delta\, \7\Xi_2\Big]=0.
\label{B9} \eea We now proceed as before and apply the variational
derivative with respect to $\7A^{*\mu}_A$. This yields: \bea
0=\frac{\delta}{\delta \7A^{*\mu}_A}\Big( \7\gamma\,\Big[ \dd{\hat
S_1}{\defpar}-\7\delta\, \7\Xi_2\Big] \Big)
=-(\7\gamma\,\eta_\mu+\qcommut{\7C}{\eta_\mu})^A, \label{B12} \eea
where \bea \eta_\mu=\frac{i\theta^{\alpha\beta}}{4}\, \Big(
2\scommut{\6_\alpha\7A_\mu}{\6_\beta\7C}
-\6_\mu\scommut{\7A_\alpha}{\6_\beta\7C}
-\qcommut{\7A_\mu}{\scommut{\7A_\alpha}{\6_\beta\7C}} \Big).
\label{B11} \eea The decomposition of $\eta_\mu$ reads thus \bea
&&\eta_\mu=\eta_\mu^{[2]}+\eta_\mu^{[3]}, \label{B13}
\\
&&\eta_\mu^{[2]}=\frac{i\theta^{\alpha\beta}}{4}\,
(2\scommut{\6_\alpha\7A_\mu}{\6_\beta\7C}
-\6_\mu\scommut{\7A_\alpha}{\6_\beta\7C}), \label{B14}
\\
&&\eta_\mu^{[3]}=-\frac{i\theta^{\alpha\beta}}{4}\,
\qcommut{\7A_\mu}{\scommut{\7A_\alpha}{\6_\beta\7C}}. \label{B15}
\eea At second and third order in the fields Eq.\ (\ref{B12})
yields: \bea \hgz\,\eta_\mu^{[2]}=0, \quad
\hgz\,\eta_\mu^{[3]}+\hgo\,\eta_\mu^{[2]}+\qcommut{\7C}{\eta_\mu^{[2]}}=0.
\label{B16} \eea Again, there exists $\xi_\mu^{[2]}$ such that
$\eta_\mu^{[2]}=\hgz\, \xi_\mu^{[2]}$, and a convenient choice
turns out to be \bea \xi^{[2]}_\mu=\frac{i\theta^{\alpha\beta}}{4}
\scommut{\d_\alpha \hat A_\mu+\hat F^{[0]}_{\alpha\mu}}{\hat
A_\beta}. \eea Inserting this result in the second equation, the
latter reads $\hgz\,(\eta_\mu^{[3]}-\hgo\,\xi_\mu^{[2]}
-\qcommut{\7C}{\xi_\mu^{[2]}})=0$. Again, one concludes that there
exists $\xi^{[3]}_\mu$ such that
\begin{equation}
\eta_\mu^{[3]}-\hgo\,\xi_\mu^{[2]}-\qcommut{\7C}{\xi_\mu^{[2]}}
=\hgz\,\xi_\mu^{[3]}\,,
\end{equation}
a convenient choice being \bea \xi^{[3]}_\mu=
\frac{i\theta^{\alpha\beta}}{4} \scommut{\qcommut{\hat
A_\alpha}{\hat A_\mu}}{\hat A_\beta}\,. \eea The resultant
$\xi_\mu^{[3]}$ satisfies
$\hgo\xi_\mu^{[3]}=-\qcommut{\7C}{\xi_\mu^{[3]}}$ which implies
$\hgo\int d^nx\  \7A^{*\mu}_A\, \xi_\mu^{[3]A}=0$. All in all this
yields \bea \dd{\hat S_1}{\defpar}-\7\delta\, \7\Xi_2=\7\gamma\,
\7\Xi_1 \label{B17} \eea with $\7\Xi_1=-\int d^nx\  \7A^{*\mu}_A\,
(\xi_\mu^{[2]}+\xi_\mu^{[3]})^A$. Explicitly one obtains \bea
\7\Xi_1=-\frac{i\theta^{\alpha\beta}}{4}\int d^nx\ \7A^{*
\mu}_A\,\scommut{\7F_{\alpha\mu}+\d_\alpha \7A_\mu}{\7A_\beta}^A.
\label{B18} \eea Using (\ref{B8}) and (\ref{B17}) in (\ref{ordd}),
we obtain
\begin{equation}
\dd{\hat S}{\defpar}=\hat B_0+\7s\, \7\Xi
\end{equation}
with $\hat\Xi=\hat\Xi_1+\hat\Xi_2$ and $\hat B_0= \dd{\hat
S_0}{\defpar}-\7\delta\, \7\Xi_1$.

We now turn to the question on how to ``invert''
$\hat\gamma^{[0]}$. In the space of fields and their derivatives,
we introduce the following new coordinates $\7y^\alpha$,
$\7z^\alpha$, $\7w^i$:
\begin{equation}
  \begin{aligned}
\{\7y^\alpha\}&=\{\partial_{(\nu_1}\dots\partial_{\nu_l}
\7A^A_{\mu)}\}\,,\quad \{\7z^\alpha\}=
\{\partial_{(\nu_1}\dots\partial_{\nu_l}  \6_{\mu)}  \7C^A\},
\\
\{\7w^i\}&= \{\7C^A,\7\chi^{[0]A}_\Delta\},\
\7\chi^{[0]A}_\Delta\equiv \6_{(\nu_1}\dots \6_{\nu_{l}}
\7F^{{[0]}A}_{\mu)\lambda} \}, \label{hatvar}
  \end{aligned}
\end{equation}
where $l=0,1,\dots$ and $\7F^{{[0]}A}_{\mu\nu}=\partial_\mu\hat
A^A_\nu- \partial_\nu\hat A^A_\mu$. These variables are
independent and complete in the sense that every local function of
fields and their derivatives can again be uniquely expressed in
terms of them.

We now define the contracting homotopy
\begin{equation}
\label{A31}\hat\rho^{[0]} f(\hat y,\hat z,\hat
w)=\int_0^1\frac{dt}{t}[ \hat y^{\alpha}\frac{\partial^L
}{\partial \hat z^\alpha}f] (t\hat y,t\hat z,\hat w),
\end{equation}
which satisfies
\begin{equation}
\label{A333} \{\hat\gamma^{[0]},\hat\rho^{[0]}\}f(\hat y,\hat
z,\hat w)= f(\hat y,\hat z,\hat w)-f(0,0,\hat w).
\end{equation}
It follows that any $\hgz$-closed function that vanishes when
$\hat y=\hat z=0$, is $\hgz$-exact. This is the case for $\eta$,
$\eta_\mu^{[2]}$, and
$\eta_\mu^{[3]}-\hgo\,\xi_\mu^{[2]}-\qcommut{\7C}{\xi_\mu^{[2]}}$
because all of them depend explicitly on $\hat z^\alpha$. In
particular, $\hgz \eta =0$ implies $\eta = \hgz \xi^\prime$ with
$\xi^\prime=\hat\rho^{[0]}\eta$.

However, the expression for $\xi^\prime=\hat\rho^{[0]}\eta$ is
rather complicated because of higher derivatives of $C$ in $\eta$.
Because the general solution to $\eta = \hgz \tilde\xi$ is
$\tilde\xi=\xi^\prime+\nu$ with $\hgz\nu=0$, one can use this
freedom to arrive at the particular solution $\xi$ given above. A
way to get the expressions for $\xi$, $\xi^{[2]}_\mu$, and
$\xi^{[3]}_\mu$ used in the text is to consider $\hat\rho^{[0]}_*$
which coincides with $\hat\rho^{[0]}$ on the variables $\hat
y,\hat z, \hat w$ but satisfies Leibnitz rule on the star
polynomials $\eta$, $\eta_\mu^{[2]}$, and
$\eta_\mu^{[3]}-\hgo\,\xi_\mu^{[2]}-\qcommut{\7C}{\xi_\mu^{[2]}}$
in these variables.

\subsection*{Appendix C: Proof of proposition
\bref{prop:comm-sol-gen}} \label{ccc}
\def\theequation{C.\arabic{equation}}
\setcounter{equation}{0}

\addcontentsline{toc}{subsection}{Appendix C: Proof of proposition
\bref{prop:comm-sol-gen}}

Given arbitrary $\Lambda$ and  $r$, together with $W_\mu$
satisfying~\eqref{eq:MC} and \eqref{eq:wmu-eq}, it can be directly
checked that $f^c_\mu$ and $h^c$ defined by \eqref{eq:f1}
and~\eqref{eq:f2} satisfy~\eqref{eq:comm-1} and \eqref{eq:comm-2}.

Conversely, given $f^c_\mu$ and $h^c$ satisfying~\eqref{eq:comm-1}
and \eqref{eq:comm-2} and the boundary
conditions~\eqref{eq:bound}, $\Lambda$, $r$, and $W_\mu$ can be
constructed order by order. Taking $\st{0}{r}=h^c|_\U-C$ and
$\st{0}{\Lambda}=1$, the equations~\eqref{eq:f1} and \eqref{eq:f2}
are satisfied at order zero in $\defpar$. Here $h^c|_\U$ denotes
the function obtained by putting to zero all variables except for
the undifferentiated ghosts $C^A$. Because $\gamma$ commutes with
putting to zero these variables, equation~\eqref{eq:comm-1}
implies that $\st{0}{r}$ satisfies~\eqref{eq:MC}. Notice also that
$(A_\mu+W_\mu) + (A_\mu^A+W^A_\mu)\dd{\st{0}{r}}{C^A}$ satisfies
equation~\eqref{eq:comm-2} for any $\bar\gamma$-closed $W_\mu$
provided $\st{0}{r}$ satisfies~\eqref{eq:MC}.

Assume that we have found $\st{k}{\Lambda}$, $\st{k}{r}$, and
$\st{k}{W}_\mu$ satisfying~\eqref{eq:MC} and \eqref{eq:wmu-eq}
such that
\begin{align}
\label{eq:kth}
h^c&=\st{k}{\Lambda}^{-1}(C+\st{k}{r})\st{k}{\Lambda}+\st{k}{\Lambda}^{-1}
\gamma\st{k}{\Lambda}+\defpar^{k+1}\sigma^{(k+1)}+O(\defpar^{k+2})\,,\\
  f^c&=\st{k}{\Lambda}^{-1}((A_\mu+\st{k}{W}_\mu) + (A_\mu^A+\st{k}{W}{}^A_\mu)\dd{\st{k}{r}}{C^A})\st{k}{\Lambda}
+\st{k}{\Lambda}^{-1}\gamma\st{k}{\Lambda}+\defpar^{k+1}\sigma_\mu^{(k+1)}+O(\defpar^{k+2})\,.
\end{align}
It then follows from equations~\eqref{eq:comm-1} and
\eqref{eq:comm-2} that $\sigma^{(k+1)}$ and $\sigma_\mu^{(k+1)}$
satisfy
\begin{equation}
\label{covp}\bar\gamma \sigma^{(k+1)}=0\,, \qquad \bar\gamma
\sigma^{(k+1)}_\mu=D_\mu \sigma^{(k+1)}\,.
\end{equation}
The general solution $\sigma^{(k+1)}$ is given by
\begin{equation}
\sigma^{(k+1)}=\bar\gamma
\lambda^{(k+1)}+r^{(k+1)}\,,\label{sigmak}
\end{equation}
where $r^{(k+1)}=C^A (r^{(k+1)})_A^B T_B$  satisfies
\begin{equation}
\bar\gamma r^{(k+1)}=0\,.
\end{equation}
This follows from the covariant Poincar\'e lemma for generic Lie
algebras proved in~\cite{Barkallil:2002fp} (see also
\cite{Brandt:1990gy,Dubois-Violette:1992ye} for the reductive
case). Indeed,
$\sigma^{(k+1)}=j^{(k+1)}(\chi,C)+\bar\gamma\mu^{(k+1)}$, where
the $\chi$ variables have been defined in \eqref{3.33}. By
considering the terms which depend only on $\chi$ and $C$, the
second equation of \eqref{covp} reduces to
\begin{equation}
\frac{\partial j^{(k+1)}}{\partial x^\mu}+D_\mu
\chi^A_\Delta\frac{\partial j^{(k+1)}}{\partial \chi^A_\Delta}+
\bar\gamma l^{(k+1)}(\chi,C)=0.\label{chiceq}
\end{equation}
In this equation, consider a  decomposition according to the
homogeneity in the original fields and assume (without loss of
generality) that $j^{(k+1)}$ starts at order M, with $\chi$
replaced by the abelian $\chi^{[0]}$'s defined in appendix {\bf
B}. To lowest order, \eqref{chiceq} coincides with the left hand side of eq.
(4.57) of ~\cite{Barkallil:2002fp} (in the particular case of form
degree 0 and k=1; the fact that the index is upper instead of
lower and that $\bar \gamma$ differs from $\gamma^R_1$ by a sign
in the first term plays no role in the following argument.)
According to the first paragraph after (4.57), we can deduce that
$j^{(k+1)}_M=\delta^1_M r^{(k+1)}+\bar\gamma
n^{(k+1)}_M(\chi^{[0]},C)$, where $ r^{(k+1)}$ depends only on the
undifferentiated ghosts, $r^{(k+1)}=C^B(r^{(k+1)})^A_BT_A$ with
$(r^{(k+1)})^A_B$ constants. Applying $\bar\gamma$, we get
$\bar\gamma r^{(k+1)}=0$. Now the abelian $\chi^{[0]}$'s can be
completed to non abelian ones to deduce that
$j^{(k+1)}=r^{(k+1)}+\bar\gamma n^{(k+1)}(\chi,C)$ with
$\bar\gamma r^{(k+1)}=0$.

For $\sigma^{(k+1)}$ given by \eqref{sigmak}, the general solution
for $\sigma^{(k+1)}_\mu$ is
\begin{equation}
\sigma^{(k+1)}_\mu = D_\mu \lambda^{(k+1)}+w^{(k+1)}_\mu +A^A_\mu
\dl{C^A}r^{(k+1)}\,,
\end{equation}
with $w^{(k+1)}_\mu=w^{(k+1)}_\mu[A,\defpar]$ satisfying
$\bar\gamma w^{(k+1)}_\mu=0$.

Taking then
\begin{equation}
  \begin{gathered}
    \st{k+1}{\Lambda}={\rm exp}({\defpar^{k+1} \lambda^{(k+1)}})\st{k}{\Lambda}
    \,, \qquad \st{k+1}{r}=\Big(\st{k+1}{\Lambda} h^c
      \st{k+1}{\Lambda}^{-1} - \, C\Big)\Big|_{\U}\,,\\
      {\st{k+1}{W}}_\mu=\st{k}{W}_\mu+\defpar^{k+1}w^{k+1}_\mu\,,
\end{gathered}
\end{equation}
one finds that equations~\eqref{eq:comm-1} and \eqref{eq:comm-2}
hold up to order $k+2$, provided that $\st{k+1}{r}-\st{k}{r}$
starts at order $\defpar^{k+1}$. This is indeed the case as one
can see by putting to zero all variables except for the
undifferentiated ghosts in equation~\eqref{eq:kth} and expressing
$\st{k}{r}$ through $h^c|_{\U}$.

It remains to be shown that $\st{k+1}{r}$
satisfies~\eqref{eq:MC}. Indeed, by putting to zero all variables
except for the undifferentiated ghosts in
equation~\eqref{eq:comm-1}, one concludes that $h^c|_{\U}$
satisfies~\eqref{eq:comm-1} and then that
$\st{k+1}{r}=(\st{k+1}{\Lambda}h^c\st{k+1}{\Lambda}^{-1})|_{\U}-C$
satisfies~\eqref{eq:MC}.

\subsection*{Appendix D: Proof of theorem~\bref{thm:amb}}

\def\theequation{D.\arabic{equation}}
\setcounter{equation}{0}

\addcontentsline{toc}{subsection}{Appendix D: Proof of
theorem~\bref{thm:amb}}

Combining proposition \bref{lem5} with proposition
\bref{prop:comm-sol-gen}, the ambiguity in the SW map is expressed
as a composition of the ``commutative'' gauge ambiguity described
by $\Lambda$ and the substitutions ${C^A\to C^A+C^Br^A_B,\,\,
A^A_\mu\to A^A_\mu  +W^A_\mu+(A^B_\mu+W^B_\mu)r^A_B}$. The only
nontrivial point in the proof is to show that the commutative
gauge ambiguity is described as a noncommutative gauge ambiguity
with some $\hat\Lambda$.

Given a particular solution $f_\mu,\g$ to \eqref{3.29} and
\eqref{3.30} and given
\begin{equation}
\hat\Lambda(t)= P\exp_*(\int_0^td\tau\hat\lambda(\tau))\,,\quad
\frac{d}{dt}\hat\Lambda(t)= \hat\Lambda(t)*\hat\lambda(t)\,,\quad
\hat\Lambda(0)={\mathbf 1}\,,
\end{equation} with arbitrary
$\hat\lambda(t)=\hat\lambda^A[A;\defpar,t]\,T_A$, let us introduce
the one parameter family of solutions
\begin{equation}
\label{eq:comm-turned}
  \begin{aligned}
  {h}^\prime(t)&={\hat\Lambda}^{-1}(t)*h*\hat\Lambda(t)+
  {\hat\Lambda}^{-1}(t)*\gamma \hat\Lambda(t)\,,\\
  f_\mu^\prime(t)&={\hat\Lambda}^{-1}(t)*f_\mu*\hat\Lambda(t)+
  {\hat\Lambda}^{-1}(t)*\d_\mu \hat\Lambda(t)\,.
  \end{aligned}
\end{equation}
It is the unique solution to the differential equations
\begin{equation}
\label{eq:diff-eq} \dd{{h}^\prime(t)}{t}=\gamma
\hat\lambda(t)+\qcommut{{h}^\prime(t)}{\hat\lambda(t)}\, \qquad
\dd{f_\mu^\prime(t)}{t}=\d_\mu\hat\lambda(t)+\qcommut{f_\mu^\prime(t)}{\hat\lambda(t)}\,,
\end{equation}
subject to the boundary conditions ${h}^\prime (0)=h$ and
$f_\mu^\prime(0)=f_\mu$. Similar equations hold for $f^c_\mu,\g^c$
and ${f_\mu^c}^\prime(t),{h^c}^\prime(t)$ with star multiplication
replaced by standard multiplication.

Using \eqref{eq:4.1} with $f^c_\mu,\g^c$ replaced by
${f_\mu^c}^\prime(t),{h^c}^\prime(t)$, one gets
\begin{multline}
\label{eq:der-1}
  \dd{{h^\prime(t)}}{t}=\dd{h}{A_{\mu,{(\sigma)}}}\Big|_{A_\nu \, \to
    \,{f^c_\nu}^\prime(t),\,\, C \,\to \,{h^c}^\prime(t)}
  \d_{{(\sigma)}}(\d_\mu \lambda+\commut{{f^c_\mu}^\prime(t)}{\lambda})
\,+
\\
+\, \dd{h}{C_{,(\sigma)}}\Big|_{A_\nu \, \to
    \,{f^c_\nu}^\prime(t),\,\, C \,\to \,{h^c}^\prime(t)}
\d_{(\sigma)}(\gamma \lambda
  +\commut{{h^{c}}^\prime(t)}{\lambda})
\,=
\\
=\, \left(\dd{h}{A_{\mu,{(\sigma)}}}
  \d_{{(\sigma)}}(\d_\mu \tilde\lambda+\commut{A_\mu}{\tilde\lambda})
+\dd{h}{C_{,(\sigma)}}\d_{(\sigma)}(\gamma \tilde\lambda
  +\commut{C}{\tilde\lambda})\right)\Big|_{A_\nu \, \to
    \,{f^c_\nu}^\prime(t),\,\, C \,\to \,{h^c}^\prime(t)}
\end{multline}
and
\begin{multline}
\label{eq:der-2}
    \dd{f^\prime_\mu(t)}{t}=\dd{f_\mu}{A_{\nu,{(\sigma)}}}\Big|_{A_\rho
    \to {f^c_\rho}^\prime(t)}\d_{(\sigma)}(\d_\nu \lambda +
  \commut{{f^c_\nu}^\prime(t)}{\lambda})
\,=
\\
=\, \left(\dd{f_\mu}{A_{\nu,{(\sigma)}}}\d_{(\sigma)}(\d_\nu
\tilde\lambda +
  \commut{A_\nu}{\tilde\lambda})\right)\Big|_{A_\rho
    \to {f^c_\rho}^\prime(t)}\,,
\end{multline}
where
$\tilde\lambda[{f^c}^\prime[A;\defpar,t],\defpar]=\lambda[A;\defpar]$
and the repeated multi-index $(\sigma)$ denotes a summation over
the derivatives of the fields as in \eqref{eq:comm-gamma}.
Applying $\tilde\lambda_{,(\sigma)}\dl{C_{,(\sigma)}}$ to
\eqref{3.29} and \eqref{3.30} and using the result in the right
hand sides of the equations~\eqref{eq:der-1} and \eqref{eq:der-2}
one arrives at \eqref{eq:diff-eq} with
\begin{equation}
\hat\lambda[A;\defpar,t] =\lambda^A_{,(\sigma)}
\left(\dd{{h^c}^\prime}{C^A_{,(\sigma)}}\Big|_{A_\mu\,\to\,{f^c_\mu}^\prime[A;\defpar,t]}\right)\,.
\end{equation}

\section*{Appendix E: Proof of
proposition~\bref{prop:nondeg-symm}}
\def\theequation{E.\arabic{equation}}
\setcounter{equation}{0}

\addcontentsline{toc}{subsection}{Appendix E: Proof of
proposition~\bref{prop:nondeg-symm}}

For $J_\mu^\nu=\theta_{\mu\rho}\eta^{\rho\nu}$,
equations~\eqref{eq:eta-theta-pres} imply
\begin{equation}
  J^\mu_\rho \lambda^\rho_\nu - J^\rho_\nu \lambda_\rho^\mu=0\,.
\end{equation}
We first solve this condition and then describe those $\lambda$
that also preserve $\eta$ (and hence $\theta$). Let
$\eta_{\mu\nu}$ and $\theta_{\mu\nu}$ be the components of the
symmetric, respectively antisymmetric, non-degenerate bilinear
forms $\eta$ and $\theta$ on $V$, the space of space-time vectors.
The defining relation for $J$ then reads
\begin{equation}
\theta(a,b)=\eta(Ja,b)\,.
\end{equation}
Since $J$ cannot in general be diagonalized over $\mathbb R$, it
is useful to consider first $K=J^2$, which is a symmetric operator
with respect to $\eta$. Because we have assumed $K$ to be
diagonalizable over $\mathbb R$, we have
\begin{equation}
K=\left(
\begin{array}{cccc}
K_1     & 0      &\ldots  & 0\\
0       &K_2     &\ldots  & 0\\
\ldots  & \ldots &\ldots   &\ldots\\
0       & 0      &\ldots   & K_N
\end{array}
\right)\,, \qquad K_\alpha=t_\alpha {\mathbf 1}_\alpha\,.
\end{equation}
where $t_\alpha$ are different eigenvalues of $K$, $t_\alpha \ne
t_\beta$ for $\alpha \ne \beta$, no summation over $\alpha$ is
implied, and ${\mathbf 1}_\alpha$ is the $m_\alpha \times
m_\alpha$ unit matrix. We denote by $N$ the number of different
eigenvalues of $K$ so that $m_1+\ldots+m_N=n$.

It follows from $\commut{J}{K}=0$ that $J$ has also a block
diagonal structure in this basis:
\begin{equation}
J=\left(
\begin{array}{cccc}
J_1     & 0      &\ldots  & 0\\
0       &J_2     &\ldots  & 0\\
\ldots  & \ldots &\ldots   &\ldots\\
0       & 0      &\ldots   & J_N
\end{array}
\right)\,, \qquad K_\alpha=J_\alpha^2\,.
\end{equation}
Similarly, the bilinear form $\eta$ is also block diagonal, with
blocks denoted by $\eta_\alpha$.

The condition $\commut{\lambda}{J}=\commut{\lambda}{K}=0$ now
implies that $\lambda$ is also block diagonal, with blocks
$\lambda_\alpha$. The problem then decomposes into
\begin{equation}
\label{eq:alpha-sym}
 \commut{\lambda_\alpha}{J_\alpha}=0\,, \qquad
 \eta_\alpha(\lambda_\alpha a,b)
+ \eta_\alpha( a,\lambda_\alpha b)=0\,,
\end{equation}
for any $a,b$ belonging to the eigenspace $V_\alpha$ of $K$ with
eigenvalue $t_\alpha$. If $\lieg_\alpha$ is the Lie algebra of
$\lambda_\alpha$ satisfying \eqref{eq:alpha-sym}, the Lie algebra
$\lieg$ of solutions to \eqref{eq:eta-theta-pres} is isomorphic to
the direct sum
\begin{equation}
  \lieg\simeq\lieg_1 \oplus \ldots \oplus \lieg_N\,.
\end{equation}

When analyzing equations \eqref{eq:alpha-sym} for a given
$\alpha$, one can always assume that $J_\alpha^2=\pm \id_\alpha$.

\begin{itemize}

\item[$J_\alpha^2=-\id_\alpha$:] In this case, the eigenspace has even
dimension $2n_\alpha$ and $J_\alpha$ can be interpreted as a
complex structure. The condition that $\eta_\alpha(J_\alpha a,b)=-
\eta_\alpha(a,J_\alpha b)\,\,\, \forall a,b\in V_\alpha$ implies
that $\eta_\alpha(J_\alpha a,J_\alpha b)=\eta_\alpha(a,b)\,\,\,
\forall a,b\in V_\alpha$ which in turn means that $\eta_\alpha$ is
(pseudo)hermitian. Equations \eqref{eq:alpha-sym} are then
respectively the conditions that $\lambda_\alpha$ is complex
linear and that it preserves the hermitian form $\eta_\alpha$.
Thus $\lambda_\alpha$ belongs to the unitary algebra
$u(n^+_\alpha,n^-_\alpha),$ $n^+_\alpha+n^-_\alpha=n_\alpha$,
where $(n^+_\alpha,n^-_\alpha)$ is the signature of the hermitian
form $\eta_\alpha$.

\item[$J_\alpha^2=\id_\alpha$:] In this case, it is useful to interpret
$J_\alpha$ as a so-called ``polarization structure''. For such a
$J_\alpha=\id_\alpha$, one can always introduce a basis of
eigenvectors $e_i,\bar e_{\bar i}$ corresponding to the
eigenvalues $+ 1$ and $-1$, respectively. It then follows from
$\eta_\alpha(J_\alpha e_i,e_j)+\eta_\alpha(e_i,J_\alpha e_j)=0$
that $\eta_\alpha(e_i,e_j)=0$. The same arguments show that
$\eta_\alpha(\bar e_{\bar i}, \bar e_{\bar j})=0$. Finally,
non-degeneracy of $\eta_\alpha$ implies that the number of $e_i$
and $\bar e_{\bar i}$ coincides. In the basis $e,\bar e$ it is
easy to solve the equations \eqref{eq:alpha-sym}. Indeed, a
general solution to $\commut{\lambda_\alpha}{J_\alpha}=0$ is given
by a block diagonal matrix $\lambda_\alpha$ with two arbitrary
blocks of dimension $n_\alpha \times n_\alpha$. At the same time
the second equation in \eqref{eq:alpha-sym} says that only one of
them is independent. Thus the Lie algebra of solutions
$\lambda_\alpha$ is $gl(n_\alpha,{\mathbb R})$.
\end{itemize}

\addcontentsline{toc}{section}{Bibliography}



\providecommand{\href}[2]{#2}\begingroup\raggedright\endgroup

\end{document}